\documentclass[a4paper,11pt]{article}

\usepackage[applemac]{inputenc}
\usepackage[english]{babel}

\usepackage{amsmath}
\usepackage{amssymb}
\usepackage{amsfonts}
\usepackage{units}
\usepackage{color}
\usepackage{cite, mcite}
\usepackage{xy}
\usepackage{slashed}
\usepackage{parskip}
\xyoption{all}

\usepackage{graphicx}
\usepackage{a4wide}

\newcommand{\tr}{{\rm{tr}}}

\newcommand{\cG}{\mathcal{G}}

\newcommand{\cL}{\mathcal{L}}

\newcommand{\cN}{\mathcal{N}}
\newcommand{\cO}{\mathcal{O}}
\newcommand{\cP}{\mathcal{P}}

\newcommand{\cS}{\mathcal{S}}
\newcommand{\cV}{\mathcal{V}}

\def\sst#1{{\scriptscriptstyle #1}}
\def\oneone{\rlap 1\mkern4mu{\rm l}}
\def\CP{{{\mathbb C}{\mathbb P}}}
\def\0{{\sst{(0)}}}
\def\1{{\sst{(1)}}}
\def\2{{\sst{(2)}}}
\def\3{{\sst{(3)}}}
\def\4{{\sst{(4)}}}
\def\5{{\sst{(5)}}}
\def\6{{\sst{(6)}}}
\def\7{{\sst{(7)}}}
\def\8{{\sst{(8)}}}

\def\crampest{\medmuskip = 1mu plus 1mu minus 1mu}
\def\uncramp{\medmuskip = 4mu plus 2mu minus 4mu}

\def\be{\begin{equation}}
\def\ee{\end{equation}}
\def\ba{\begin{array}}
\def\ea{\end{array}}
\def\bea{\begin{eqnarray}}
\def\eea{\end{eqnarray}}
\def\nn{\nonumber}
\newcommand{\ft}[2]{{\textstyle\frac{#1}{#2}}}
\def\fft#1#2{{\frac{#1}{#2}}}

\def\cG{{{\cal G}}}
\def\im{{\rm i\,}}

\newcommand{\eq}[1]{(\ref{#1})}
\newcommand{\w}[1]{\\[0.#1cm]}

\makeatletter
\@addtoreset{equation}{section}
\makeatother

\begin{document}


\begin{titlepage}

\begin{flushright}
MIFPA-14-30

\end{flushright}

\bigskip

\begin{center}

\vskip 2cm

{\LARGE \bf Correlation Functions in \\[4mm] $\omega$-Deformed  $\cN = 6$
Supergravity} \\

\vskip 1cm

{\bf A.~Borghese$^1$, Y.~Pang$^1$, C.N.~Pope$^{1,2}$ and E.~Sezgin$^1$} \\

\vskip 1cm

{\em
$^1$ {\it George P. \& Cynthia Woods Mitchell  Institute
for Fundamental Physics and Astronomy,\\
Texas A\&M University, College Station, TX 77843, USA }\\

\vskip .4cm

$^2${\it DAMTP, Centre for Mathematical Sciences,
 Cambridge University,\\  Wilberforce Road, Cambridge CB3 OWA, UK}
} \\

\end{center}

\vskip 2cm

\begin{center} {\bf ABSTRACT} \\[3ex]

\begin{minipage}{13cm}
\small

Gauged $\cN=8$ supergravity in four dimensions is now known to admit a
deformation characterized by a real parameter $\omega$ lying in the interval
$0\le\omega\le \pi/8$. We analyse
the fluctuations about its anti-de Sitter vacuum, and show
that the full $\cN=8$
supersymmetry can be maintained by the boundary conditions only for $\omega=0$.
For  non-vanishing $\omega$,  and requiring that there be no
propagating spin $s>1$ fields
on the boundary, we show that $ \cN=3$ is the maximum degree of supersymmetry
that can be preserved by the boundary conditions.  We then construct
in detail the consistent
truncation of the  $\cN=8$ theory to give $\omega$-deformed $SO(6)$
gauged $\cN=6$
supergravity, again with $\omega$ in the range
$0\le\omega\le \pi/8$.  We show that this theory admits fully $\cN=6$
supersymmetry-preserving
boundary conditions not only for $\omega=0$, but also for $\omega=\pi/8$.
These two theories are related by a $U(1)$ electric-magnetic duality.
We observe that the only
three-point functions that depend on $\omega$ involve the coupling
of an $SO(6)$ gauge field with the $U(1)$ gauge field and a scalar or
pseudo-scalar field. We compute these correlation functions and compare
them with those of the undeformed $\cN=6$ theory. We find that
 the correlation functions in the $\omega=\pi/8$ theory holographically
correspond
 to amplitudes in the $U(N)_k\times U(N)_{-k}$ ABJM model in which
the $U(1)$ Noether current is
replaced by a dynamical $U(1)$ gauge field.  We also show that
the $\omega$-deformed $\cN=6$ gauged supergravities can be obtained
via consistent reductions from the eleven-dimensional or ten-dimensional type
IIA supergravities.

\end{minipage}

\end{center}

\vspace{2cm}

\vfill

\end{titlepage}

\setcounter{page}{1}

\tableofcontents


\section{Introduction}

  For thirty years after its construction in 1982 \cite{deWit:1982ig},
the $SO(8)$ gauged maximally supersymmetric
$\cN=8$ supergravity was widely considered to be a unique theory.
Interestingly, using the embedding tensor formulation \cite{deWit:2007mt},
it was recently realized that there is a one-parameter
extension of the theory, commonly denoted by $\omega$, associated with  
a mixing of the electric and magnetic vector fields employed in
the $SO(8)$ gauging \cite{DalLAgata:2012bb,dewitnicnew}.  Inequivalent
$\cN=8$ theories are parameterised by $\omega$ in the range
$0\le\omega\le \pi/8$.  This development
has raised numerous interesting
questions, such as its possible higher-dimensional string/M theory origin
and the consequences of the $\omega$ deformation for the
holographic dual theory.

 In this paper, as a step towards addressing the holography-related
questions in
particular, we shall begin by showing that if we retain all the fields
of the $SO(8)$ gauged supergravity, then the maximum degree of
supersymmetry that is compatible with any consistent boundary conditions in the
$\omega$-deformed $SO(8)$ theory is $\cN=3$.  A key assumption in reaching this
conclusion is that we allow
only Dirichlet boundary
conditions for the bulk fields with  spins $s>1$, since Neumann boundary
conditions would give rise to associated propagating spin $s>1$ fields
in the
dual boundary field theory.  We then show that if we truncate the
$\cN=8$ theory to $\cN=6$, the resulting theory still has a
non-trivial $\omega$ deformation
parameter, with $0\le\omega\le\pi/8$, and for two specific
inequivalent choices of the $\omega$ parameter, namely $\omega=0$ or
$\omega=\pi/8$,
it is possible to impose boundary conditions that are
compatible with the full $\cN=6$ supersymmetry.
These two theories are related to each by a $U(1)$ electric-magnetic duality.

We
construct the full bosonic Lagrangian and supersymmetry transformation of
$\omega$-deformed $SO(6)$ gauged supergravity as a consistent truncation of
the $\omega$-deformed $\cN=8$ theory, generalising similar results for the
undeformed $SO(6)$ gauged $\cN=6$  supergravity \cite{ferrara}. We compute the
three-point correlation functions
of the theory at tree-level, focusing on those which depend on the
value of the $\omega$ parameter.
We find that  the only such three-point functions involve the coupling of
the $SO(6)$ gauge fields with the $U(1)$ gauge field and a scalar or
pseudo-scalar field.  (The $\cN=6$ supergravity has $SO(6)\times U(1)$ 
gauge fields, with the scalars and fermions being charged under $SO(6)$ but
not under the $U(1)$.) We also compute these correlation functions in the
undeformed
$\cN=6$ theory. In comparing these results, and finding their possible
holographic interpretation,
make use of Witten's observation \cite{Witten:2003ya} that an
electric-magnetic duality rotation in the bulk
corresponds to a so called $S$-transformation of the boundary CFT,
in which a global $U(1)$ symmetry is gauged and an off-diagonal Chern-Simons
term is introduced. Interestingly, the $U(N)_k\times U(N)_{-k}$
 ABJM model already contains the desired $U(1)\times U(1)$ sector. Thus,
we suggest 
that the holographic dual of the $\omega=\pi/8$ theory is not a new
CFT, as it would be in Witten's generic framework, but is instead the
ABJM model itself, in the sense that the processes
involving the Noether current $J$ and those involving the
dynamical $U(1)$ in the ABJM model are described by ostensibly distinct
bulk theories with $\omega=0$ and $\omega=\pi/8$ respectively.  The 
precise relationship involves the interchange in the CFT correlation
functions of a $U(1)$ Noether
current and a topological current already present in the ABJM model.

  We also give a discussion of the embedding of the $\omega$-deformed
theories into higher dimensions.  Unlike the $\omega$-deformed
$\cN=8$ supergravities, whose embedding into eleven dimensions could
be expected to involve the introduction of the ``dual graviton'' in
$D=11$ \cite{dewitnicnew}, we find that the $\omega$-deformed
$\cN=6$ gauged supergravities can be embedded into the standard
eleven-dimensional or ten-dimensional type IIA supergravities.
The essential reason is that the $\omega$-deformed $\cN=6$ theories
are equivalent, after making an appropriate $U(1)$ duality rotation,
at the level of the equations of motion, and since no fields have minimal
couplings to the $U(1)$ gauge potential there is no obstruction to
performing the necessary dualisation.   Furthermore, the consistent
Kaluza-Klein sphere reductions always operate at the level of the
equations of motion; one cannot write a sphere-reduction ansatz that
can be substituted into the higher-dimensional action. For this reason,
the embedding of the $\omega$-deformed $\cN=6$ theories can be
implemented by making the appropriate $U(1)$ duality rotation on the
usual embedding ansatz.  In the case of an embedding into eleven dimensions
there would still be a non-local relation between the ans\"atze for
inequivalent values of $\omega$, since the bare $U(1)$ gauge potential
appears in the eleven-dimensional metric ansatz.  If one instead
considers the embedding into ten-dimensional type IIA supergravity,
where the $U(1)$ field comes from the Ramond-Ramond 2-form, the bare $U(1)$
gauge potential appears nowhere in the reduction ansatz, and so the
embeddings for different values of $\omega$ can be locally related.

The plan of this paper is as follows.  In section 2 we begin by reviewing
some of the key features of the $\omega$-deformed $\cN=8$ gauged
supergravities.  We then construct an expansion, up to the first few
orders in fields, around the maximally-symmetric $\cN=8$ AdS$_4$ vacuum, with
the object of identifying the leading-order interaction terms in which the
effect of the $\omega$ parameter becomes apparent.  We find that this
occurs in the trilinear couplings between an $SO(6)$ gauge field, a
$U(1)$ gauge field, and a scalar or pseudoscalar field.  We also
set up the Fefferman-Graham (FG) expansions for all the fields in the
AdS$_4$ background.  In section 3, we present in detail the
consistent reduction of the $\omega$-deformed $\cN=8$ gauged supergravities
to $\cN=6$.  We also show that the $\omega$ parameter remains non-trivial,
in the sense that it parameterises theories related by a $U(1)$ duality
rotation that lies outside the $SO^*(12)$ global symmetry group of the theory.
In section 4 we study the supersymmetry transformations of the
boundary fields in the FG expansion around AdS$_4$, and we show that the
full $\cN=6$ supersymmetry is preserved not only in the undeformed
$\omega=0$ theory but also in the $\omega=\pi/8$ theory.  Section 5
contains detailed calculations of the 3-point amplitudes associated
with the $\omega$-dependent trilinear couplings identified in section 2.
These calculations would also have wider applicability in other 
situations where one has gauge fields obeying Neumann boundary 
conditions as as well as gauge fields obeying Dirichlet boundary conditions.
We discuss the interpretation of these results in the holographic dual
boundary theory in section 6.  The paper ends with conclusions in section 7.
In appendix A we discuss the embedding of the $\omega$-deformed $\cN=6$
theories in eleven and ten dimensions, and in appendix B we discuss
supersymmetric boundary conditions for the boundary fields in the
$\omega$-deformed $\cN=8$ gauged supergravities.

\section{$\omega$-Deformed $\cN=8$ Gauged Supergravity}

   In this section we begin by reviewing some of the key aspects of the
construction of the $\omega$-deformed $\cN=8$ gauged supergravities.
The, for later convenience, we present the first few terms in an expansion
of the Lagrangian order-by-order in powers of the fields.  Finally in
this section, we study the details of the Fefferman-Graham expansions for
the linearised solutions around the $\cN=8$ supersymmetric AdS$_4$
background of the $\omega$-deformed theory.

\subsection{Review of the $\omega$-deformed theory}

The $\cN=8$ supergravity multiplet consists of the fields
\be
(e_\mu^a, \psi_\mu^i, A_\mu^{IJ}, \chi^{ijk}, \phi^{ijk\ell})\ ,
\ee
where $e_\mu^a$ is the vielbein, $\psi_\mu^i$ are Weyl gravitini ($i=1,...,8$),   $A_\mu^{IJ} = A_\mu^{[IJ]}$ are the vector fields ($I,J=1,...,8$),   $\chi^{ijk}= \chi^{[ijk]}$ are the spin $1/2$ Weyl fermions and
\be
\phi^{ijk\ell}= \phi^{[ijk\ell]} = (\phi_{ijk\ell})^\star = \ft{1}{4!}\epsilon^{ijk\ell mnpq} \phi_{mnpq}\label{su8duality}
\ee
are the scalar fields, which parameterize the
coset $E_{7(7)}/SU(8)$. In the
56-dimensional representation, an element of $E_{7(7)}$ can be written as
\begin{align}
\cV = \left(\begin{matrix}  u_{ij}{}^{IJ} &  v_{ij KL} \\
& \\
v^{k\ell IJ} & u^{k\ell}{}_{KL} \end{matrix} \right)\ ,
\label{56bein}
\end{align}
where $u^{ij}{}_{IJ}= (u_{ij}{}^{IJ})^\star$ and $v_{ijIJ}= (v^{ijIJ})^\star$ and
\be
\cV^\star = \theta\cV\theta\ ,\qquad \cV^\dagger \Omega \cV = \Omega\ ,\qquad
\theta=\left( \begin{matrix} 0& \oneone_{28}\\[1mm] \oneone_{28} & 0 \end{matrix} \right)\ ,\qquad \Omega
=\left(\begin{matrix} \oneone_{28} & 0\\[1mm] 0& -\oneone_{28} \end{matrix}
\right)\ .\label{nutheta}
\ee
The $56$-bein $\cV$ transforms by right-multiplication under a
rigid $E_{7(7)}$ and by left-multiplication under a local $SU(8)$. Thus
the indices $[ij]$ and $[k\ell]$ are local $SU(8)$ indices,
whilst $[IJ]$ and $[KL]$ are rigid $E_{7(7)}$ indices. The standard $SO(8)$
gauged $\cN=8$ supergravity theory uses the $56$-bein defined above.
To obtain the $\omega$-deformed version the theory, it suffices to
perform the scalings \cite{dewitnicnew}
\be
u_{ij}{}^{IJ} \to e^{i\omega} u_{ij}{}^{IJ}\ ,\qquad v_{ijIJ}
\to e^{-i\omega} v_{ijIJ}\ ,
\ee
with  $\omega \in (0,\pi/8]$. The bosonic sector of the resulting $\omega$-deformed theory is described by the Lagrangian
\bea
\cL_{\rm bos}  &= & \cL_{\rm EH} + \cL_{\rm gauge}+ \cL_{scalar} - eV
\nn\\[2mm]
&= &  \tfrac{1}{2} \, e \, R - \tfrac18 \, e \, \left[ \im
F^+_{\mu \nu \, IJ}  \left(2S^{IJ,KL}-\delta^{IJ}\delta^{KL}\right)  F^{+\mu \nu}{}_{KL} + \mbox{h.c.} \right]
- \tfrac{1}{96} \, e \, P_{\mu}^{ijk\ell} \, P^{\mu}_{ijk\ell}
\nn\\[1mm]
&& - e \, g^{2} \, ( - \tfrac{3}{4} \, A_{ij} \, A^{ij} + \tfrac{1}{24} \, A_i{}^{jk\ell} \, A^{i}{}_{jk\ell} ) \ ,
\label{bL}
\eea
where $\cL_{\rm gauge}$ and  $\cL_{scalar}$  are the kinetic terms for
the gauge and scalar fields, $V$ is the potential,  and
$S^{IJ,KL}$ is a function
of the scalar fields defined by
\be
\im \left(u^{ij}{}_{IJ} +v^{ij IJ}\right) S^{IJ,KL} = u^{ij}{}_{IJ} \ .
\label{defS}
\ee
It can be shown that $S^{IJ,KL}=S^{KL,IJ}$. Further definitions are
\be
\begin{split}
F_{\mu \nu}{}^{IJ}  &= 2 \, \partial_{[\mu} A_{\nu]}{}^{IJ} - 2 \, g \, A_{[\mu}{}^{IM} A_{\nu]}{}^{MJ} \ ,
\\[0.2cm]
P_{\mu}^{ijk\ell}  &=  -2{\sqrt 2} \, \left[ u^{ij}{}_{IJ} \, D_{\mu}(A) v^{k\ell IJ} - v^{ij IJ} \, D_{\mu}(A)  u^{k\ell}{}_{IJ} \right]\ ,
\\[0.2cm]
D_{\mu} (A) u_{ij}{}^{IJ} &=  \partial_{\mu} u_{ij}{}^{IJ} -2 \, g \, A_{\mu}{}^{M[I} \, u_{ij}{}^{J]M} \ .
\end{split}
\label{3defs}
\ee
The $SU(8)$ tensors built out of scalar fields are defined as
\be
A^{ij} = \tfrac{4}{21}  \, T_{k}{}^{ikj} \quad , \qquad A_i{}^{jk\ell} = - \, \tfrac{4}{3} \, T_{i}{}^{[jk\ell]} \ ,
\ee
where
\be
T_{i}{}^{jkl} = \left(e^{-i \omega} \, u^{kl}{}_{IJ} + e^{i \omega} \, v^{kl IJ}) \, (u_{im}{}^{JK} \, u^{jm}{}_{KI} - v_{im JK} \, v^{jm KI}\right) \ ,
\ee
The local supersymmetry transformations, neglecting terms of cubic or
higher order in the fermions, are given by \cite{Lu:2014fpa}
\bea
\delta e_\mu{}^a &=& {\bar\epsilon}^i \gamma^a\psi_{\mu i} + \mbox{h.c.}\ ,
\nn\\[1mm]
\delta\psi_\mu^i  &=& 2 D_\mu\epsilon^i +\frac{1}{2\sqrt 2} H^-_{\rho\sigma}{}^{ij} \gamma^{\rho\sigma} \gamma_\mu\epsilon_j
+ {\sqrt 2} g A^{ij}\gamma_\mu \epsilon_j\ ,
\nn\\[1mm]
\delta A_\mu{}^{IJ} &=& -\left(e^{i\omega} u_{ij}{}^{IJ} + e^{-i\omega} v_{ij IJ} \right) \left( {\bar\epsilon}_k\gamma_\mu \chi^{ijk}
+ 2{\sqrt 2} {\bar\epsilon}^i\psi_\mu{}^j \right)+ \mbox{h.c.}\ ,
\nn\\[1mm]
\delta\chi^{ijk} &=& -P_\mu{}^{ijk\ell}\gamma^\mu\epsilon_\ell + \tfrac32 \gamma^{\mu\nu} H^-_{\mu\nu}{}^{[ij}\epsilon^{k]}
-2g A_\ell{}^{ijk} \epsilon^\ell   \ ,
\nn\\[1mm]
\left( \delta \cV_M{}^{ij} \right) \cV^{M k\ell}  &=&  2\sqrt{2} \left( \bar\epsilon^{[i}\chi^{jk\ell]}+\tfrac1{24}\varepsilon^{ijk\ell mnpq}\,
\bar\epsilon_m \chi_{npq}\right)\ ,
\eea
where $\cV_M{}^{k\ell} = \left(u^{k\ell}{}_{IJ},v^{k\ell IJ}\right)$, and
\be
H^-_{\mu\nu}{}^{ij} =  \left(  e^{-i\omega} u^{ij}{}_{IJ} F_{1\mu\nu}{}^{IJ} + e^{i\omega} v^{ij IJ} F^-_{2\mu\nu}{}^{IJ}\right) \ ,
\ee
where
\bea
F^+_{1\mu\nu IJ} &=& \tfrac12 \left( \im
         G^+_{\mu\nu IJ} + F^+_{\mu\nu IJ} \right) \ ,
\qquad
F^+_{2\mu\nu}{}^{IJ} = \tfrac12 \left( \im
    G^+_{\mu\nu IJ} - F^+_{\mu\nu IJ} \right) \ ,
\label{def1}\w2
G^{+\mu\nu}{}_{IJ}  &=& \frac{4\im}{e} \frac{\delta\cL_{\rm gauge}}{
           \delta F^+_{\mu\nu IJ}}\ .
\label{def2}
\eea
The covariant derivative of $\epsilon^i$ is given by
\bea
D_\mu \epsilon^i &=& \partial_\mu \epsilon^i -\tfrac14 \omega_\mu{}^{ab} \gamma_{ab} \epsilon^i +\tfrac12 Q_\mu{}^i{}_j \epsilon^j\ ,
\nn\\[0.2cm]
Q_\mu{}^i{}_j  &=& \tfrac23 \left[ u^{ik}{}_{IJ} D_\mu(A) u_{jk}{}^{IJ} - v^{ik IJ} D_\mu (A) v_{jk IJ} \right]\ .
\eea
It is useful to note that the field equations for the vector fields,
up to fermionic terms which can be handled straightforwardly 
\cite{dewitnicnew}, and the Bianchi identities for their field strengths, can together be
expressed as the $56$-dimensional vector equation
\be
\partial_\mu \left[ e \left( \begin{matrix}  F_1^{+\mu\nu} \\[1mm]  F_2^{+\mu\nu}\end{matrix}  \right)
+e\theta \left( \begin{matrix}  F_1^{+\mu\nu} \\[1mm]
F_2^{+\mu\nu}\end{matrix} \right)^\star \right]=0\ ,
\label{eom}
\ee
where $\theta$ is defined in (\ref{nutheta}).
In the limit of vanishing $SO(8)$ coupling $g$ \cite{Cremmer:1979up}, this equation is
invariant under a rigid $E_{7(7)}$ symmetry.  This equation serves to
define $H+_{\mu\nu}$
as well as $G^+_{\mu\nu}$, the latter taking the form
\be
\im \, (e^{-\, i \omega}u^{ij}{}_{IJ} + e^{ \, i \omega} \, v^{ij IJ} ) \, G^+_{\mu \nu \, IJ} = (e^{-\, i \omega}u^{ij}{}_{IJ} - e^{ \, i \omega} \, v^{ij IJ} ) \, F^{+}_{\mu \nu \, IJ} \ .
\ee

The ostensible  $Sp(56)$ symmetry of \eq{eom} for $g=0=\omega$ is broken
down to $E_7(7)$ by the requirement of the consistency of the transformations
that rotate $F_{\mu\nu}$ and $G_{\mu\nu}$ into each other with the equation
(\ref{def2}).  This is a stringent condition, since \eq{def2} shows that
$G_{\mu\nu}$  is not an independent field but rather a functional of
$F_{\mu\nu} $ and the scalar fields.  The fact that consistency is
achieved by for the  $E_{7(7)}$ symmetry is seen manifestly from the
observation that the following relation holds:
\be
\cV \left( \begin{matrix} F^+_{1\mu\nu} \\[1mm] F^+_{2\mu\nu} \end{matrix} \right) =
\left( \begin{matrix} H^+_{\mu\nu} \\[1mm] 0 \end{matrix} \right)\ .
\label{duality1}
\ee
This equation is manifestly $E_{7(7)}$ invariant for vanishing $SO(8)$ coupling constant, and vanishing $\omega$.
The introduction of $\omega$-dependent phase factors takes $\cV$
outside $E_{7(7)}$ but it is still inside $Sp(56)$, in such
a way  that the theory is still locally supersymmetric.  The
further turning on of the $\omega$ parameter is consistent with the
local $SO(8)$ and with supersymmetry.

In the following section, where we shall consider the consistent
truncation of the
$\cN=8$ theory, we shall need the identities
\bea
0 &=&
\left( u^{k\ell}{}_{IJ} + v^{k\ell IJ} \right) \left( u_{ij}{}^{JK}\,u^{mn}{}_{KI} - v_{ij JK}\, v^{mnKI} \right) - \tfrac23 \delta^{[m}_{[i} T_{j]}{}^{n]k\ell}\ ,
\label{e1}\w2
0 &=& \tfrac43 T_i{}^{jk\ell} +A_{2i}{}^{jk\ell} +2 A_1^{j[k} \delta_i^{\ell]}\ ,
\label{e2}\w2
0 &=& 18 A_1^{ik} A_{1kj} - A^i_{2k\ell m} A_{2j}{}^{k\ell m} - \mbox{trace} \ .
\label{e3}
\eea
We shall also need the relation \cite{deWit:2002vz}
\be
Q^{[k}_{\mu\  [i} \delta^{\ell]}{}_{j]}  = u_{ij}{}^{IJ} D_\mu(A) u^{k\ell}{}_{IJ} -v_{ij IJ} D_\mu(A) v^{k\ell IJ}\ .
\label{defQ}
\ee
Finally, we record a convenient parametrization of the $E_{7(7)}/SU(8)$
coset element in the so-called symmetric gauge, in which it takes the form
\be
\cV = \exp \left[ \begin{matrix} 0 & -\tfrac{1}{2\sqrt 2} \phi_{ijkl} \\[1mm] -\tfrac{1}{2\sqrt 2} \phi^{ijkl} & 0 \end{matrix} \right] \ .
\label{cV}
\ee
In this gauge the $I,J$ indices are no longer distinguishable from the
$i,j$ indices.

\subsection{Expansion of the $\omega$-deformed $\cN=8$ supergravity around
maximally supersymmetric AdS$_4$} \label{couplings}

In what follows, we will use the following abbreviations
\begin{align} \label{abbreviation 1}
(\phi \cdot \bar{\phi})^{ij}{}_{kl} \equiv \phi^{ijmn} \, \phi_{mnkl} \quad , \qquad (\phi \cdot \bar{\phi} \cdot \phi)^{ij , \, kl} \equiv \phi^{ijmn} \, \phi_{mnpq} \, \phi^{pqkl}, \,\quad a=-\frac1{2\sqrt{2}} \ .
\end{align}
Using the coset representative in symmetric gauge as given in \eq{cV},  $u$ and $v$ up to fourth order we expand
\bea
u^{ij}{}_{IJ} & = & \delta^{ij}{}_{IJ} + \tfrac{a^2}{2} \, (\phi \cdot \bar{\phi})^{ij}{}_{IJ} + \tfrac{a^4}{4!} \, (\phi \cdot \bar{\phi} \cdot \phi \cdot \bar{\phi})^{ij}{}_{IJ} + \cO(\phi^{6})\ ,
\nn\w2
v^{ijIJ}  &=&  \, a \, \phi^{ijIJ} + \tfrac{a^3}{3!} \, (\phi \cdot \bar{\phi} \cdot \phi)^{ij , \, IJ} + \cO(\phi^{5}) \ .
\eea
The expansion of $P_\mu^{ijk\ell}$ up to cubic order in fields is
\bea
P_{\mu}^{ijkl} &=&  \partial_{\mu} \phi^{ijkl}  + 4 \, g \, \phi^{I[ijk} \, A_{\mu}{}^{l]J} \, \delta_{IJ} 
\nn\w2
& & + \tfrac{a^2}{3} [ (\phi \cdot \bar{\phi})^{[ij}{}_{IJ} \, \partial_{\mu} \phi^{kl]IJ} - \phi^{ijMN} \, \phi^{klPQ} \, \partial_{\mu} \phi_{MNPQ} ] \ .
\eea
In order to derive the gauge kinetic terms we have to solve order by order in $\cG^{+}_{\mu \nu \, IJ}$ equation \eq{def2}. The result is the following
\bea
i \, G^{+}_{\mu \nu \, IJ} &=&  F^{+(1)}_{\mu \nu \, IJ} + F^{+(2)}_{\mu \nu \, IJ} - 2 \, a \, e^{2 i \omega} \phi^{IJ KL} \, F^{+(1)}_{\mu \nu \, KL}
\nn\w2
&& -  2 \, a \, e^{2 i \omega} \, \phi^{IJKL} \, F^{+(2)}_{\mu \nu \, KL}
+ 2 \, a^{2} \, e^{2 i \omega} \, \phi^{IJMN} \, \phi^{MNKL} F^{+(1)}_{\mu \nu \, KL} +\cdots
\label{expandF}
\eea
where we have denoted with $F^{(1)}$ and $F^{(2)}$ the abelian and nonabelian part of the field strengths respectively
\be
F^{(1)}_{\mu \nu}{}^{IJ} = 2 \, \partial_{[\mu} A_{\nu]}{}^{IJ} \ ,
\qquad F^{(2)}_{\mu \nu}{}^{IJ} = - 2 \, g \, A_{[\mu}{}^{IM} A_{\nu]}{}^{MJ} \ .
\ee
Using the above lemmata, we find
\bea
A^{ij} & = & + e^{- \im \omega} \, \delta^{ij} + e^{- \im \omega} \, \delta^{ij} \, \tfrac{a^{2}}{24} \, | \phi |^{2}
+ e^{\im \omega} \, \tfrac{a^{3}}{3} \, (\phi \cdot \bar{\phi} \cdot \phi)^{i I, \, j J} \, \delta_{IJ}
\nn\w2
&& + e^{- \im \omega} \, \big[ \delta^{ij} \, \tfrac{a^{4}}{864} \, (|\phi|^{2})^{2}
- \tfrac{a^{4}}{6} \, (\phi \cdot \bar{\phi})^{Ki}{}_{LI} \, (\phi \cdot \bar{\phi})^{Lj}{}_{KJ} \, \delta^{IJ} \big]
\nn\w2
&& +e^{ \im \omega}\, \big[ -\ft{19a^5}{420}(\phi \cdot \bar{\phi} \cdot \phi\cdot \bar{\phi}\cdot\phi)^{iI,jI}
+\ft{a^5}{63}|\phi|^{2}(\phi \cdot \bar{\phi} \cdot \phi)^{iI,jI}
\nn\w2
&& \qquad+\ft{a^5}{21}(\phi \cdot \bar{\phi})^{i\ell}_{~~KI}(\phi \cdot \bar{\phi})^{JK}_{~~~k\ell}\phi^{kjIJ}\big]\ .
\label{A2Exp}
\eea
\bea
A_{i}{}^{jkl}  & = & - e^{\im \omega} \, 2 \, a \, \delta_{iI} \, \phi^{Ijkl}
- e^{-\im \omega} \, 3 \, a^{2} \, (\phi \cdot \bar{\phi})^{[jk}{}_{iI} \, \delta^{l]I}
\nn\w2
&& - e^{\im \, \omega} \, \big[ \tfrac{a^{3}}{9} \, \delta_{iI} \, \phi^{Ijkl} \, |\phi|^{2}
+ 2 \, a^{3} \, \phi^{KI[jk} \, \phi_{Ki mn} \, \phi^{l] J mn} \, \delta^{IJ} \big]
\nn\w2
&& +e^{-\im \, \omega} \,\Big[\ft{a^4}{12} (\phi \cdot \bar{\phi}\cdot\phi\cdot\bar{\phi})^{[jk}_{~~~iI}\delta^{\ell]}_I
-\ft{7a^4}{72}|\phi|^{2}(\phi \cdot \bar{\phi})^{[jk}_{~~~iI}\delta^{\ell]}_I
\nn\w2
&& \qquad +\ft{a^4}3(\phi \cdot \bar{\phi})^{m[j}_{~~~nI}\delta^k_I(\phi \cdot \bar{\phi})^{\ell]n}_{~~~im}-a^4(\phi \cdot \bar{\phi})^{[jk}_{~~~IJ}(\phi \cdot \bar{\phi})^{\ell]I}_{~~~iJ}\Big]\ .
\label{A4Exp}
\eea
In obtaining the above results, we have used the following properties, related to self-duality of the scalar fields
\bea
\phi^{rst m} \, \phi_{rst i}  &= & + \tfrac{1}{8} \, \delta_{i}{}^{m} \, | \phi |^{2} \ ,
\nn\w2
\phi^{r mnp} \, \phi_{r ijk} & = & - \tfrac{1}{16} \, \delta_{ijk}{}^{mnp} \, | \phi |^{2}
+ \tfrac{9}{4} \, \delta_{[i}{}^{[m} \, (\phi \cdot \bar{\phi})^{np]}{}_{jk]} \ .
\eea
Note that in \eq{A2Exp} and \eq{A4Exp} the even and odd powers of scalar fields are multiplied by different $\omega$ phases. When computing the scalar potential we take the T tensor components times their complex conjugates and the $\omega$ dependence drops out in the terms which contain an even number of scalar fields. An $\omega$ dependent phase shows up in the other terms, containing an odd number of scalar fields. This is consistent with the fact that the value of the cosmological constant and the scalar spectrum (obtained from the order zero and order two terms) are insensitive to the $\omega$-deformation. Finally notice that, if we want to compute the scalar potential up to fourth order in the scalar fields we only need to compute $A_{i}{}^{jkl}$ up to third order, due to the absence of the zeroth order contribution to this T-tensor component.

Next, we expand the bosonic Lagrangian (\ref{bL}) to fourth order in excitations around the maximally supersymmetric AdS$_{4}$ vacuum. The vacuum solution corresponds to $ \phi_{ijkl} = 0 = A_{\mu}{}^{IJ}$. Using the formula given above, we find
\bea
e^{-1} \, \cL_{\rm scalar} &=& - \tfrac{1}{96} \, \partial_{\mu} \phi^{ijkl} \, \partial^{\mu} \phi_{ijkl}
+ \tfrac{1}{24} \, g \, (\partial_{\mu} \phi^{ijkl}) \, \phi_{Iijk}  \delta_{lJ} \, A^{\mu \, IJ} + \text{c.c.}
\nn\w2
&&  - \tfrac{a^{2}}{144} \, (\partial_{\mu} \phi^{ijmn}) (\partial^{\mu} \phi_{mnkl}) \, (\phi \cdot \bar{\phi})^{ij}{}_{kl}
\nn\w2
&& - \tfrac{a^{2}}{286} \left[ \phi^{ijmn} \, \phi^{klpq} \, (\partial_{\mu} \phi_{ijkl}) (\partial^{\mu} \phi_{mnpq}) + \text{c.c} \right]
\nn\w2
&& - \tfrac{1}{24} \, g^{2} \, [ \tfrac{1}{8} \, | \phi |^{2} \, A_{\mu}{}^{IJ} \, A^{\mu \, IJ}
+ 3 \, (\phi \cdot \bar{\phi})^{IK}{}_{JL} \, A_{\mu}{}^{IJ} \, A^{\mu \, KL}] \ .
\eea
The expansion of the gauge kinetic terms, on the other hand, up to fourth order in field, yields
\bea
e^{-1} \cL_{\rm gauge} &=& - \tfrac{1}{2} \, \partial_{[\mu} A_{\nu]}{}^{IJ} \, \partial^{\mu} A^{\nu \, IJ}
 + g \, \partial_{[\mu} A_{\nu]}{}^{IJ} \, A^{\mu \, IM} A^{\nu \, MJ}
\nn\w2
&& + {\rm Re} \{ a e^{2 i \omega} \phi^{IJKL} \} \, \partial_{[\mu} A_{\nu]}{}^{IJ} \, \partial^{\mu} A^{\nu \, KL}
\nn\\
&&
+\tfrac{1}{2} \, {\rm Im}\{ a e^{2 i \omega} \phi^{IJKL} \} \, \epsilon^{\mu \nu \rho \sigma} 
\partial_{\mu} A_{\nu}{}^{IJ} \, \partial_{\rho} A_{\sigma}{}^{KL} \, +
\nn\w2
&& - {\rm Re} \{ a^{2} e^{2 i \omega} (\phi \cdot \phi)^{IJ , \, KL} \} \, \partial_{[\mu} A_{\nu]}{}^{IJ} \, \partial^{\mu} A^{\nu \, KL}
\nn\\
&&
- \tfrac{1}{2} \, {\rm Im} \{ a^{2} e^{2 i \omega} (\phi \cdot \phi)^{IJ , \, KL} \} \,
\epsilon^{\mu \nu \rho \sigma} \, \partial_{\mu} A_{\nu}{}^{IJ} \, \partial_{\rho} A_{\sigma}{}^{KL}
\nn\w2
&&  - 2 \, g \, {\rm Re} \{ a \, e^{2 i \omega} \, \phi^{IJKL} \} \, \partial_{[\mu} A_{\nu]}{}^{IJ} A^{\mu \, KM} A^{\nu \, ML}
\nn\\
&&
- \, g \, {\rm Im}\{ a \, e^{2 i \omega} \, \phi^{IJKL} \} \, \epsilon^{\mu \nu \rho \sigma} \,
\partial_{\mu} A_{\nu}{}^{IJ} A_{\rho}{}^{KM} A_{\sigma}{}^{ML}
\nn\w2
&& - \tfrac{1}{2} \, g^{2} \, A_{[\mu}{}^{IM} A_{\nu]}{}^{MJ} A^{\mu \, IN} A^{\nu \, NJ} \ .
\eea
Finally,  the potential up to fifth order in scalar fields, is given by
\bea
V  &= & - 6 - \tfrac{a^{2}}{3} \, |\phi|^{2} + \tfrac{a^{4}}{216} \, (|\phi|^{2})^{2} - \tfrac{a^{4}}{3} \, (\phi \cdot \bar{\phi})^{Ki}{}_{LI} \, (\phi \cdot \bar{\phi})^{Li}{}_{KI} 
\nn\w2
&& -\Big[ \ft{a^5}{10}e^{\im 2\omega}(\phi \cdot \bar{\phi}\cdot\phi\cdot\bar{\phi}\cdot\phi)^{iI,iI}
-\ft{a^5}{36}e^{\im 2\omega}|\phi|^{2}(\phi \cdot \bar{\phi}\cdot\phi)^{iI,iI}+c.c\Big]\ .
\eea
Note that this is independent of the deformation parameter $\omega$.  In fact,
as we argue in section 3, the $\omega$ parameter does not enter at all
in the scalar potential $V$ of the $\cN=6$ theory.

\subsection{Fefferman-Graham expansions in the linearized
 $\omega$-deformed $\cN=8$ supergravity}

Here we study the variation of the boundary fields arising in the
Fefferman-Graham (FG) expansion, in which certain convenient gauge choices
are made for the fields of spins $s\ge 1$.  Since we shall be using
the symmetric gauge for the representative of the the coset
$E_{7(7)}/SU(8)$, there will be no distinction between $SU(8)$ and
$SO(8)$ indices.  Therefore, in this subsection we shall
use $I,J,K,...$ to denote $SO(8)$ indices, which then allows us
to use the indices $i,j,k,..$ to label the coordinates of the
three-dimensional
boundary of AdS$_4$. For the fermionic fields, the avoidance of
$SU(8)$ indices will be facilitated by going over to Majorana basis.
In doing so, we shall, for convenience, include $\omega$-dependent
phases as follows:
\be
e^{\ft{i}{2}\omega}\epsilon_L + e^{-\ft{i}{2}\omega}\epsilon_R\ \   \to \epsilon^I  \ \ \mbox{(Majorana)}\ ,
\qquad  e^{-\ft{i}{2}\omega}\chi_R+ e^{\ft{i}{2}\omega}\chi_L \ \ \to\ \ \chi^{IJK}\ \ \mbox{(Majorana)}\ ,
\ee
where the $SU(8)$ indices on the chiral spinors are suppressed.
It will also prove to be convenient to define the real and imaginary parts
of the scalar fields as
\be
\phi^{IJKL} = \cS^{IJKL}+\im\cP^{IJKL}\ .
\ee
It follows from (\ref{su8duality}) that
\be
\cS^{IJKL}= \fft1{4!}\,\epsilon^{IJKLMNPQ}\, \cS_{MNPQ}\ ,\qquad \cP^{IJKL}= -\fft1{4!}\,\epsilon^{IJKLMNPQ}\, \cP_{MNPQ}\ .
\label{su8duality2}
\ee

Choosing the gauges
\be
e^{\hat{0}}_{0} = \frac1{z_0}\ , \qquad e^{\hat{0}}_{i}=0\ , \qquad e^{\hat{r}}_{0}=0\ , \qquad A^{IJ}_{0} = 0\ ,
\qquad \psi_{0}^I = 0\ ,
\label{gaugechoices}
\ee
the equations of motion then determine the falloff behaviour of the fields to be
\bea
e^{\hat{r}}_{i} &=& \frac1{z_0}(e^{\hat{r}}_{(0)i}+z_0^2i\,
 e^{\hat{r}}_{(2)_i}+z_0^3\, e^{\hat{r}}_{(3)i}+\cdots)\ ,
\nn\w2
A^{IJ}_{i} &=& A_{(0)i}^{IJ}+ z_0\, A^{IJ}_{(1)i}+\cdots \ ,
\nn\w2
{\cal S}^{IJKL}&=& z_0\, {\cal S}^{IJKL}_{(1)}
    +z_0^2 \, {\cal S}^{IJKL}_{(2)} +\cdots\ ,
\nn\w2
{\cal P}^{IJKL} &=& z_0\, {\cal P}^{IJKL}_{(1)}
     +z_0^2\, {\cal P}^{IJKL}_{(2)}+\cdots\ ,
\nn\w2
\psi^{I}_{i} &=& z_0^{-\ft12} \psi^{I}_{(0)i+}+z_0^{\ft12} \psi^{I}_{(2)i-}+z_0^{\ft32} \psi^{I}_{(3)i}+\cdots\ ,
\nn\w2
\chi^{IJK} &=& z_0^{\ft32} \chi^{IJK}_{+}+z_0^{\ft32} \chi^{IJK}_{-}  +\cdots\ ,
\label{FGexp0}
\eea
The asymptotic Killing spinor can be expressed as
\be
\epsilon^{I}=z_0^{-\ft12} \epsilon^{I}_+ + z^{\ft12}_0 \epsilon^{I}_- + z^{\ft32}_0 \epsilon^{I}_{(3)} + \cdots\ .
\ee

   Plugging the FG expansions of various fields into the supersymmetry
transformation rules, we can extract the supersymmetry variations of the
coefficients in the expansions.  Firstly, we see that
\bea
\delta e^{\hat{r}}_{(0)i}&=&\bar{{\epsilon}}^{I}_+\gamma^{\hat{r}}_{(0)}{\psi}^{I}_{(0)i+}\ ,
\nn\w2
\delta{\psi}^{I}_{(0)i+}&=&\ft12{\cal K}^{ab}_{(0)i}\gamma_{ab}{\epsilon}^{I}_++\sqrt{2}A^{~IJ}_{(0)i }{\epsilon}^{J}_+\ ,
\eea
where ${\cal K}^{ab}_{(0)i}$ is the super-torsion constructed from
${\psi}^{I}_{(0)i+}$, and we have used the fact that
$\partial_i\epsilon_+=\ft12\gamma_i\epsilon_-$.  (See, for example,
\cite{lupoto} for the explicit solution for the Killing spinors in
AdS$_{d+1}$ in Poincar\'e coordinates.)
The supersymmetry variation of the boundary data of the spin-1 field
is given as
\bea
\delta A^{IJ}_{(0)i}&=&-\Big(\cos2\omega\ \, {\epsilon}^{K}_+\gamma_{(0)i}{\chi}^{IJK}_{+}+{\rm i}\sin2\omega\ \, {\epsilon}^{K}_+\gamma_{(0)i}\gamma_5{\chi}^{IJK}_{-}\Big)+ \cdots\ ,
\nn\w2
\delta A^{IJ}_{(1)i}&=&\Big[-{\cal S}_{(1)}^{IJKL}{\epsilon}^{M}_+\gamma_{(0)i}{\chi}^{KLM}_{+}
-{\rm i}{\cal P}_{(1)}^{IJKL}{\epsilon}^{M}_+\gamma_{(0)i}\gamma_5{\chi}^{KLM}_{-}-2\sqrt{2}{\epsilon}^{[I}_+{\psi}^{J]}_{i(3)-}
\nn\w2
&&+D_{i}(A_{(0)}) \left( \cos2\omega\ \, {\epsilon}^K_{+}{\chi}^{IJK}_{-}
+{\rm i}\sin2\omega\ \, {\epsilon}^K_{+}\gamma_5{\chi}^{IJK}_+ \right) \Big] +\cdots \ .
\eea
where the ellipses refer to term depending on ${\psi}^{r}_{(0)i+}$,
which vanish for the Dirichlet boundary conditions that we shall impose on
the gravitini in the next section when we analyze  the
supersymmetry-preserving boundary conditions. There remains the supersymmetry
variation of the boundary data of the spin-${\ft12}$ and spin-$0$ fields,
which take the form
\bea
\delta{\chi}^{IJK}_{+}&=&-{\cal S}^{IJKL}_{(2)}{\epsilon}^L_++2{\rm i}{\cal P}^{IJKL}_{(1)}\gamma_5{\epsilon}^L_--{\rm i}\slashed{D}{\cal P}_{(1)}^{IJKL}\gamma_5{\epsilon}^L_+\ ,
\nn\w2
\delta{\chi}^{IJK}_{-}&=&2{\cal S}^{IJKL}_{(1)}{\epsilon}^L_--{\rm i}{\cal P}^{IJKL}_{(2)}\gamma_5{\epsilon}^L_+
+\slashed{D}{\cal S}_{(1)}^{IJKL}{\epsilon}^L_+\ ,
\nn\w2
\delta{\cal S}^{IJKL}_{(1)}&=&4\Big(\bar{{\epsilon}}^{[I}_+{\chi}_{-}^{JKL]}+\ft1{4!}\varepsilon^{IJKLMNPQ}\bar{{\epsilon}}^{M}_+{\chi}_{-}^{NPQ}\Big)\ ,
\nn\w2
\delta{\cal P}^{IJKL}_{(1)}&=&-4{\rm i}\Big(\bar{{\epsilon}}^{[I}_+\gamma_5{\chi}_{+}^{JKL]}
-\ft1{4!}\varepsilon^{IJKLMNPQ}\bar{{\epsilon}}^{M}_+\gamma_5{\chi}_{+}^{NPQ}\Big)\ ,
\nn\w2
\delta{\cal S}^{IJKL}_{(2)} &=& 4\Big(\bar{{\epsilon}}^{[I}_-{\chi}_{+}^{JKL]}+\ft1{4!}\varepsilon^{IJKLMNPQ}\bar{{\epsilon}}^{M}_-{\chi}_{+}^{NPQ}
\nn\w2
&&+\bar{{\epsilon}}^{[I}_+\slashed{D}{\chi}_{+}^{JKL]}+\ft1{4!}\varepsilon^{IJKLMNPQ}\bar{{\epsilon}}^{M}_+\slashed{D}{\chi}_{+}^{NPQ}\Big)\ ,
\nn\w2
\delta{\cal P}^{IJKL}_{(2)}&=&-4{\rm i}\Big(\bar{{\epsilon}}^{[I}_-\gamma_5{\chi}_{-}^{JKL]}-\ft1{4!}\varepsilon^{IJKLMNPQ}\bar{{\epsilon}}^{M}_-\gamma_5{\chi}_{-}^{NPQ}
\nn\w2
&&-\bar{{\epsilon}}^{[I}_+\gamma_5\slashed{D}{\chi}_{-}^{JKL]}+
\ft1{4!}\varepsilon^{IJKLMNPQ}
\bar{{\epsilon}}^{M}_+\gamma_5\slashed{D}{\chi}_{-}^{NPQ}\Big)\ .
\eea

It should be noted that several compensating transformations have been
used when deriving the above results.  The reason for this is that
the gauge choices adopted in the FG expansion are not preserved under
the supersymmetry transformations alone. It is necessary to make
certain compensating  transformations of the fields using the
diffeomorphism, local Lorentz and local $SO(8)$ symmetries, in order to
maintain the orignal gauge choices. We denote the corresponding
transformation parameters by $(\xi^{\mu}$, $\Lambda^{a}_{~b}$ and
$O^{IJ})$, respectively.
The gauge choice for the vielbein (see (\ref{gaugechoices})) is preserved
by accompanying the supersymmetry transformations with
compensating diffeomorphism and local Lorentz transformations,
whose parameters are related to the supersymmetry parameter by
\be
\xi^{\mu}=-\int dr(\bar{{\epsilon}}^I\gamma^{\hat{0}}{\psi}^I_{\hat{r}} e^{\mu \hat{r}})\ ,
\qquad \Lambda^{\hat{0}}_{~i}=-\bar{\epsilon}^I\gamma^{\hat{0}}{\psi}^I_{i} \ .
\ee
To maintain the gauge choice $A_0^{IJ}=0$ requires a compensating
$SO(8)$ transformation with parameters determined by the conditions
\be
\partial_{z_0}O^{IJ}=-\delta_{\epsilon}A^{IJ}_{0}\ .
\ee
The  $O^{IJ}$ can be solved order by order in $z_0$, with, at leading order,
\be
O^{IJ}=z_0\, O^{IJ}_{(1)}+\cdots\ ,\qquad
O^{IJ}_{(1)}=(\cos2\omega\ \, {\epsilon}_{K+}{\chi}^{IJK}_{-}
+\im\sin2\omega\ \, {\epsilon}_{K+}\gamma_5{\chi}^{IJK}_+)\ .
\ee
The compensating transformation $O^{IJ}$ explains the derivative term
in the supersymmetry variation of $A_{(1)i}^{IJ}$. To maintain the
gauge condition ${\psi}^r_{0}=0$, the ${\epsilon}^i_{(3)}$ and higher-order
coefficients in the FG expansion  of the supersymmetry transformation
parameter need to be modified,  but these do not affect the result at
the order to which we are working. The compensating $SU(8)$
transformation needed for maintaining the symmetric gauge takes the
form $\Pi^I_{~J}=\ft23 \left( u^{IM}{}_{KL}
\delta_{\epsilon}u_{JM}{}^{KL}-v^{IMKL}\delta_{\epsilon} v_{JMKL} \right)$.
It can be seen that $\Pi^{I}_{~J}\sim {\cal O}(z_0^2)$ after using the
FG expansions of the scalar fields. Therefore, $\Pi^I_{~J}$ will
not contribute to the variation of the leading falloff coefficients.

\section{Consistent Truncation to $\omega$-Deformed $SO(6)$ Gauged \\
$\cN=6$ Supergravity }
In this section, we shall construct the full bosonic Lagrangian and
supersymmetry transformations of $\omega$-deformed $SO(6)$ gauged
supergravity as a consistent truncation of the $\cN=8$ theory summarized
in the previous section.  To keep the notation simple, we shall use the
same indices  $i,j$ and $I,J$ that in the previous
section ran from 1 to 8, but now in this section they will run from 1 to 6
for the ${\cal N}=6$ theory.   The consistent
truncation of the $\cN=8$ theory is achieved by the rules
\be
A^{7I}=0\ , \qquad A^{8I}=0\ ,\qquad \phi^{IJK7}=0\ ,\qquad \phi^{IJK8}=0\ ,
\label{tr}
\ee
in the bosonic sector, and by
\be
\psi_\mu^7=\psi_\mu^8=0\ ,\qquad \chi^{IJ7}=\chi^{IJ8}=0\ ,
\ee
in the fermionic sector. It is straightforward to check that the supersymmetry transformations of
the truncated fields remain vanishing.  Introducing the notation
\bea
A_\mu^{78} &:=&  A_\mu\ ,\qquad P_\mu^{ij 78} := P_\mu^{ij}\ ,\qquad Q^{78} := Q_\mu\ ,
\nn\w2
u^{ij}{}_{78}  &:=& u^{ij}\ ,\qquad u_{78}{}^{IJ} := u^{IJ}\ ,\qquad u^{78}{}_{78} := u\ ,
\label{defsuv}
\eea
with self-explanatory similar definitions for $v^{ij}, v^{IJ}, v)$, the
truncation rules \eq{tr} lead to the following result for the full
bosonic Lagrangian of the $\omega$-deformed $SO(6)$ gauged supergravity:
\bea
\cL_{\rm bos}&=&\tfrac{1}{2} \, e \, R-\ft18 e \Big[ \im
  F^{+\, IJ}_{\mu\nu} \left( 2 S^{IJ,KL}-\delta^{IK} \delta^{JL} \right)
F^{+\mu\nu}_{KL} + 4\im F^{+}_{\mu\nu} S^{IJ} F_{IJ} ^{+\mu\nu} 
\nn\\
&& +2\im F^{+}_{\mu\nu} ( 2S-1) F^{+\mu\nu} + \hbox{h.c.} \Big]
- \tfrac{1}{96} \, e \, P_{\mu}^{ijk\ell} \, P^{\mu}_{ijk\ell} - \tfrac18 \, e \, P_{\mu}^{ij} \, P^{\mu}_{ij} - eV \  ,
\eea
where the $S$ functions are defined by the relations
\crampest
\be
\im \left[ (u+v )( u^{ij}{}_{KL} +v^{ijKL} )
   -(u^{ij} + v^{ij})( u_{KL} + v_{KL}) \right] S^{KL,IJ}
= (u+v) u^{ij}{}_{IJ} -(u^{ij}+v^{ij}) u_{IJ}\ ,
\nn
\ee
\uncramp
\be
S^{IJ} = (u+v)^{-1} \left[ u_{IJ} - (u_{KL}+v_{KL}) S^{KL,IJ}\,\right]\ ,
\ee
\be
S= (u+v)^{-1} \left[ u-(u_{IJ}+v^{IJ}) S^{IJ}\,\right]\ ,
\nn
\ee
as can be seen from \eq{defS} and the truncation conditions \eq{tr}.
The definitions \eq{3defs} hold for the $\cN=6$ theory with all the
indices restricted to run from 1 to 6, while $P_\mu^{ij}$ is defined as
\be
P_\mu^{ij} = -2{\sqrt 2} \left[ u^{ij}{}_{IJ} D_\mu (A) v^{IJ} - v^{ij IJ}
D_\mu (A) u_{IJ} \right]\ .
\ee

The scalar potential for the $\cN=6$  theory takes the form
\be
V=g^2 \left( -\, A_{1ij} \, A_1^{ij}+\ft1{18}\, A_{2i}{}^{jk\ell} \, A_2{}^{i}{}_{jk\ell}+\ft13 \, A_{2i}{}^{j} \, A_2{}^{i}{}_{j} \right)\ ,
\label{pot2}
\ee
where the functions
\be
A_1^{ij}  = \tfrac{4}{15}  \, T_{k}{}^{ikj} \ , \qquad  A_{2i}{}^{jk\ell} = - \, \tfrac{4}{3} \, T_{i}{}^{[jk\ell]}\ ,
\qquad  A_{2i}{}^{j} = - \, \tfrac{4}{3} \, T_{i}{}^{j} \ ,
\label{a}
\ee
are defined in terms of the tensors
\be
\begin{split}
T_{i}{}^{jkl} &= \ft32\left(e^{-i \omega} \, u^{kl}{}_{IJ} + e^{i \omega} \, v^{kl IJ}) \, (u_{im}{}^{JK} \, u^{jm}{}_{KI}
- v_{im JK} \, v^{jm KI}\right)-\delta^j_iS^{kl}\ ,
\w2
S^{kl} &= \ft35\left(e^{-i \omega} \, u^{kl}{}_{IJ} + e^{i \omega} \, v^{kl IJ}) \, (u_{ij}{}^{JK} \, u^{ij}{}_{KI} - v_{ij JK} \, v^{ij KI}\right)\ .
\end{split}
\label{d12}
\ee
and
\be
\begin{split}
T_{i}{}^{j} &= \ft32\left(e^{-i \omega} \, u_{IJ} + e^{i \omega} \, v^{IJ}) \, (u_{im}{}^{JK} \, u^{jm}{}_{KI} - v_{im JK} \, v^{jm KI}\right)-\delta^j_iS\ ,
\w2
S &= \ft35\left(e^{-i \omega} \, u_{IJ} + e^{i \omega} \, v^{IJ}) \, (u_{ij}{}^{JK} \, u^{ij}{}_{KI} - v_{ij JK} \, v^{ij KI}\right)\ .
\label{d34}
\end{split}
\ee
The result \eq{pot2} is obtained by restricting the free indices of of
\eq{e3} to run from 1 to 6 and then taking the trace. The relations
\eq{a} follow from \eq{e2}, while \eq{d12} and \eq{d34} follow from
\eq{e1} by the restriction of the free indices to lie in the $SO(6)$ and
$U(1)$ directions.


The local supersymmetry transformations of the $\cN=6$ theory are obtained,
up to cubic fermions, from those of the $\cN=8$ theory by applying the
consistent truncation rules \eq{tr}, and they take the form
\bea
\delta e_\mu{}^a &=& {\bar \epsilon}^i \gamma^a\psi_{\mu i} + \mbox{h.c.}\ ,
\nn\\[1mm]
\delta\psi_\mu^i  &=& 2 D_\mu\epsilon^i +\frac{1}{2\sqrt 2} H^-_{\rho\sigma}{}^{ij} \gamma^{\rho\sigma} \gamma_\mu\epsilon_j
+ {\sqrt 2} g A_1^{ij}\gamma_\mu \epsilon_j\ ,
\nn\w2
\delta A_\mu{}^{IJ} &=& - \Big[ \left(e^{i\omega} u_{ij}{}^{IJ} + e^{-i\omega} v_{ijIJ} \right) \left( {\bar\epsilon}_k\gamma_\mu \chi^{ijk}
+ 2{\sqrt 2} {\bar\epsilon}^i\psi_\mu{}^j \right)
\nn\w2
&& \qquad + 2 \left(e^{i\omega} u^{IJ} + e^{-i\omega} v_{IJ} \right) {\bar\epsilon}_k \gamma_\mu \chi^k + \mbox{h.c.}\Big] \ ,
\nn\w2
\delta A_\mu \!\!\!\! &=& 
  \!\!\!\!
-\left(e^{i\omega} u_{ij} + e^{-i\omega} v_{ij} \right) \left( {\bar\epsilon}_k\gamma_\mu \chi^{ijk}
\!+ 2{\sqrt 2} {\bar\epsilon}^i\psi_\mu{}^j \right)\!
 +  2(e^{i\omega} u+ e^{-i\omega} v) {\bar\epsilon}_k \gamma_\mu \chi^k 
+ \! \mbox{h.c.}\ ,
\nn\w2
\delta\chi^{ijk}  &=& -P_\mu{}^{ijk\ell}\gamma^\mu\epsilon_\ell + \tfrac32 \gamma^{\mu\nu} H^-_{\mu\nu}{}^{[ij}\epsilon^{k]}
-2g A_{2\ell}{}^{ijk} \epsilon^\ell    \ ,
\nn\w2
\delta\chi^i &=& -P_\mu{}^{ij}\gamma^\mu\epsilon_j + \tfrac12 \gamma^{\mu\nu} H^-_{\mu\nu}\epsilon^i
-2g A_{2j}{}^i \epsilon^j   \ ,
\nn\w2
\left( \delta \cV_M{}^{ij} \right) \cV^M &=&  \sqrt{2}  \left( \bar\epsilon^{[i}\chi^{j]}+\tfrac1{12}\varepsilon^{ijk\ell mn}\,
\bar\epsilon_k \chi_{\ell mn}\right)\ ,
\eea
where $\cV_M{}^{k\ell} = \left(u^{k\ell}{}_{IJ},v^{k\ell IJ}\right)$.
We have made the following definitions. Firstly we define the $32$-bein
\be
{\cal V}=\left(
  \begin{array}{cc}
    U & V \\
    V^\star & U^\star \\
  \end{array}
  \right)\ ,\qquad
U :=\left( \begin{array}{cc}
    u_{ij}^{~KL} & u_{ij}\\
    u^{~KL} & u \\
  \end{array}
\right)\ ,
\qquad
V :=\left( \begin{array}{cc}
    v_{ijKL} & v_{ij} \\
    v_{KL} & v \\
  \end{array}
\right)\ .
\label{cr2}
\ee
Recalling the definitions \eq{defsuv}, it  is straightforward to check that ${\cal V}$ satisfies
\be
\cV^\star =\theta \cV  \theta \ , \qquad  \cV^{\dag}\Omega \cV =\Omega\ ,
\label{Vrels}
\ee
where
\be
\theta=\left( \begin{matrix} 0& \oneone_{16}\\[1mm] \oneone_{16} & 0 \end{matrix} \right)\ ,\qquad \Omega
=\left(\begin{matrix} \oneone_{16} & 0\\[1mm] 0& -\oneone_{16} \end{matrix} \right)
\ee
The relations (\ref{Vrels}), together with the form for $\cV$ in \eq{cr2},
show that $\cV$ is an element of $SO^\star(12)$.

Next we define
\be
\begin{split}
H^-_{\mu\nu}{}^{ij} =&  \left(  e^{-i\omega} u^{ij}{}_{IJ} F_{1\mu\nu}{}^{IJ} + e^{i\omega} v^{ij IJ} F^-_{2\mu\nu}{}^{IJ} \right)
+ \left(  e^{-i\omega} u^{ij} F_{1\mu\nu} + e^{i\omega} v^{ij} F^-_{2\mu\nu} \right) \ ,
\w2
H^-_{\mu\nu} =&  \left(  e^{-i\omega} u^{ij}{}_{IJ} F_{1\mu\nu}{}^{IJ} + e^{i\omega} v^{ij IJ} F^-_{2\mu\nu}{}^{IJ} \right)
+ \left(  e^{-i\omega} u^{ij} F_{1\mu\nu} + e^{i\omega} v^{ij} F^-_{2\mu\nu} \right) \ ,
\end{split}
\label{hh}
\ee
where $F^+_{1\mu\nu IJ}$ and $F^+_{2\mu\nu IJ}$ are defined as in \eq{def1},
but with the indices now running from 1 to 6, and
\be
 F^+_{1\mu\nu}=(\im G^+_{\mu\nu}+F^{+}_{\mu\nu})\ , \qquad
 F^+_{2\mu\nu}=(\im G^+_{\mu\nu }-F^{+}_{\mu\nu})\ .
 \label{defsF1F2}
\ee
We also have the definition similar to \eq{def2}, but
now with the indices running from 1 to 6, namely
\be
G^{+\mu\nu}  = \frac{4\im}{e} \frac{\delta\cL_{\rm gauge}}{
   \delta F^+_{\mu\nu}}\ .
\label{def33}
\ee
The covariant derivative of $\epsilon^i$ is now defined as
\bea
D_\mu \epsilon^i &=& \partial_\mu \epsilon^i -\tfrac14 \omega_\mu{}^{ab} \gamma_{ab} \epsilon^i +\tfrac12 Q_\mu{}^i{}_j \epsilon^j\ ,
\nn\w2
Q_\mu{}^i{}_j  &=& -u_{\ell j}{}^{IJ} D_\mu(A) u^{i\ell}{}_{IJ} - v_{\ell j IJ} D_\mu(A) v^{i \ell IJ}
- u_{\ell j}\partial_\mu u^{i\ell} + v_{\ell j} \partial_\mu v^{i\ell}
\w2
&& + \tfrac1{10} \delta^i_j \left[ u_{k\ell}{}^{IJ} D_\mu(A) u^{k\ell}{}_{IJ} - v_{k\ell IJ} D_\mu(A) v^{k\ell IJ}
+ u_{k\ell} \partial_\mu u^{k\ell} - v_{k\ell} \partial_\mu v^{k\ell} \right]\ .
\nn
\eea

In the limit of vanishing $SO(6)$ coupling constant, the field equations
are invariant under $SO^\star(12)$ duality transformations, as can be seen
from the field equations analogous to \eq{eom}, and the fact that
\be
{\cal V}\left(
          \begin{array}{c}
            F^+_{1\mu\nu IJ} \\
            \\
            F^+_{1\mu\nu}\\
            \\
            F^{+KL}_{2\mu\nu} \\
            \\
            F^{+}_{2\mu\nu}\\
          \end{array}
        \right)=\left(
                  \begin{array}{c}
                    H^{+}_{\mu\nu ij}\\
                    \\
                    H^{+}_{\mu\nu}\\
                    \\
                    0 \\
                    \\
                    0\\
                  \end{array}
                \right).
\ee
One can show that $\hat{U}{\cal V}$ satisfies the same properties
as ${\cal V}$, provided that $\hat U$ takes the form
\be
\hat{U}=\left(
          \begin{array}{cc}
            U_{16\times 16} & 0 \\
            0 & U^*_{16\times 16} \\
          \end{array}
        \right),\quad U_{16\times 16}=\left(
                          \begin{array}{cc}
                            SU(6)_{15\times 15} & 0 \\
                           0 & 1 \\
                          \end{array}
                        \right)\times \left(
                           \begin{array}{cc}
                             e^{\im\alpha} \oneone_{15}  & 0 \\
                             0 & e^{-3\im\alpha} \\
                           \end{array}
                         \right)\ ,
\ee
where the $U(1)_R$ factor is given by
\be
\hat{V}=\left(
          \begin{array}{cc}
            V_{16\times 16} & 0 \\
            0 & V^*_{16\times 16} \\
          \end{array}
        \right),\quad V=\left(
                           \begin{array}{cc}
                             e^{\im\alpha} \oneone_{15}  & 0 \\
                             0 & e^{-3\im\alpha} \\
                           \end{array}
                         \right)\ .
\ee
Therefore $\hat{U}$ is an element of $R$-symmetry group $U(6)$ and consequently $\hat{U}{\cal V}$ also parameterises $SO^*(12)$. Note that since $\hat V$ is an element inside the $U(6)$ subgroup of $SO^*(12)$ and ${\cal V}$ is an element in the coset $SO^*(12)/U(6)$, by definition
\be
\hat V {\cal V}(\phi)={\cal V}(\phi') \hat V,\quad \phi'^{IJ}=e^{-2\im \alpha}\phi^{IJ}\ .
\ee
On the other hand, the $\omega$-deformed $\cN=6$ theory is obtained by acting with
a matrix $\hat W$ on ${\cal V}$ from the right, where
\be
\hat W=\left(
          \begin{array}{cc}
            e^{\im \omega} \oneone_{16}  & 0 \\
            0 & e^{-\im \omega} \oneone_{16}  \\
          \end{array}
        \right).
\ee
Since $\hat W \neq \hat V$, we see that the $\omega$ deformation of the $\cN=6$ theory
cannot be absorbed by means of any $U(1)_R$ transformation.

   It is, perhaps, worth emphasising that although the above argument
shows that $\omega$ is a non-trivial parameter in the $\cN=6$ theory,
it is for the slightly subtle reason that one cannot perform any local
field redefinition that would implement a duality rotation on the
$U(1)$ gauge field in the Lagrangian itself.  One could, of course,
make such a duality rotation at the level of the equations of motion.
To see this, let us consider the duality rotation
\be
\begin{pmatrix} F_{\mu\nu} \cr G_{\mu\nu}\end{pmatrix} \longrightarrow
 \begin{pmatrix} \cos\beta & \sin\beta\cr
                  -\sin\beta & \cos\beta\end{pmatrix}
 \begin{pmatrix} F_{\mu\nu} \cr  G_{\mu\nu}\end{pmatrix} \ ,
 \label{dualU1}
\ee
where $G_{\mu\nu}=4\epsilon_{\mu\nu\alpha\beta}\partial\cL/\partial F_{\alpha\beta}$, which then implies
\be
F_{1\mu\nu}^+\longrightarrow e^{-\im\beta}\, F_{1\mu\nu}\ , \qquad
F_{2\mu\nu}^+\longrightarrow e^{\im\beta}\, F_{2\mu\nu}\ .
\ee
Thus on the full set of $SO(6)\times U(1)$ gauge field strength and their duals,
this duality transformation is implemented by the matrix
\be
\hat Z= \begin{pmatrix} Z_{16\times 16} & 0\cr
                          0 & Z^*_{16\times 16}\end{pmatrix}\ , \qquad
Z_{16\times 16}= \begin{pmatrix} \oneone_{15} & 0\cr
                                  0 & e^{-\im\beta}\end{pmatrix}\ .
\ee
It is now evident that if we right-multiply $\cV$ by the matrix $\hat V\hat Z$,
with $\alpha=\omega$ and $\beta=-4\omega$, we obtain the same result as
the right-multiplication by $\hat W$ that generates the $\omega$
deformation of the $\cN=6$ theory \cite{Dall'Agata:2014ita}.   Although it might therefore appear
that the $\omega$ deformation is trivial in the $\cN=6$ theory this is
in fact not the case, since the theory at the quantum level is specified
not by its equations of motion but rather, by its Lagrangian, and in
the Lagrangian one would have to make a non-local field redefinition of
$U(1)$ gauge potential in order to implement the $\hat V\hat Z$ transformation.
In particular, since the Lagrangian defines the nature of the
correlation functions in the dual theory, the results can, and indeed do,
depend on the value of $\omega$. Nonetheless, As we shall see in section 6, there exists a relationship between the correlation
functions of the $\omega=0$ and $\omega=\pi/8$ theories. It is also worth emphasising that since
the equations of motion in the $\omega$-deformed $\cN=6$ supergravity are
independent of the $\omega$ parameter, upon the use of a duality rotation
of the $U(1)$ gauge field, this implies that the scalar potential $V$ must
be independent of $\omega$.  This accords with what we found in section 2.2,
where the scalar potential was expanded explicitly, up to and including the
fifth order in the scalar fields.

\section{Supersymmetric Boundary Conditions in the
$\omega$-Deformed ${\cal N}=6$ Theory }

In order to calculate the holographic correlation functions it is
crucial to understand the boundary conditions imposed on the fields.
In the standard AdS/CFT dictionary, the partition function of the bulk
gravity theory is equal to the partition function
of the boundary CFT. The boundary
values of the bulk fields are identified as the external sources coupling
to certain operators on the boundary. In general, the bulk fields satisfy
second-order equations and therefore, for a single field, near the
boundary of AdS there are two boundaries values associated with
different falloff rates.
Under certain circumstance, these two boundary values allow the option
of different boundary condition choices for the bulk fields. Interestingly,
imposing different boundary conditions for the bulk fields leads to
different dual boundary CFTs. As an example, in $D=d+1$-dimensional AdS
space of unit radius with the metric
\be
ds^2=\frac{dz_0^2+\sum_{i=1}^{d}dz_i^2}{z^2_0}\ ,
\ee
a scalar field with mass-squared $m^2$ behaves as
$\phi(z)\sim z_0^{\Delta_-}\phi_-(\vec z)+z_0^{\Delta_+}\phi_+(\vec z)$
when approaching the AdS boundary, where
$\Delta_{\pm}=\ft12(d\pm\sqrt{d^2+4m^2})$. It was established
in \cite{Maldacena:1997re,Gubser:1998bc, Witten:1998qj} that
for $m^2>-\ft{d^2}4+1$, there is
a unique admissible boundary condition, for which
$\phi_-(\vec z)$ is treated as the external source on the boundary CFT.
However, for $-\ft{d^2}4<m^2<-\ft{d^2}4+1$ there are additional  possible
conditions. One can impose either a Neumann boundary condition by
identifying $\phi_+$ as an external source on the boundary, or else
a mixed condition by imposing a functional relation between $\phi_-$ and
$\phi_+$.   The Neumann boundary condition leads to an alternative
quantization on the boundary CFT \cite{Klebanov:1999tb}, while the
mixed boundary condition is interpreted as deforming the CFT by
multi-trace operators \cite{Berkooz:2002ug, Witten:2001ua}.

 For bulk Abelian gauge vectors, it was shown in \cite{Ishibashi:2004wx}
that for $d=3$, 4 and 5, both the slow-falloff and the fast-falloff parts
of the vector are normalizable. Furthermore, in \cite{Marolf:2006nd}
it was suggested that if the Neumann boundary condition is adopted for
the vector field, the dual of the vector field represents a dynamical
gauge field in the CFT.  Similarly, the possibility of imposing different
boundary conditions for spin-$\ft32$ and for the graviton has been explored
in \cite{Marolf:2006nd,Compere:2008us,Amsel:2009rr }. The Neumann boundary
conditions for the gravitino and the graviton imply the existence of
dynamical gravitino or graviton fields in the boundary theory.

    In this work, we shall focus on the case where the $\omega$-deformed
${\cal N}=6$ gauged supergravity can still have a dual CFT description.
We shall, therefore, still impose standard Dirichlet boundary conditions on
the spin-$\ft32$ and spin-2 gauge fields.  With the
boundary conditions for the spin-$\ft32$ and spin-2 fields thus determined,
the supersymmetry-preserving boundary conditions for the lower-spin fields
can then be derived from the supersymmetry transformations of the
coefficient functions  associated with the large-distance expansions
of the lower-spin fields \cite{Amsel:2009rr,Hollands:2006zu,Amsel:2008iz}.

 Turning now to the specific case of the $\omega$-deformed gauged
${\cal N}=6$ supergravity, we therefore begin, as discussed above, by
imposing Dirichlet boundary condition on the spin-2 and spin-$\ft32$ fields.
 From the supersymmetry variation of the leading coefficients of
the spin-2 and spin-$\ft32$ fields \footnote{For simplicity, we work with ${\rm AdS}_4$ of unit 
radius, which corresponds to setting $g=1/\sqrt{2}$.}
\bea
\delta e^{\hat{r}}_{(0)i} &=&
 \bar{\epsilon}^{I}_+\gamma^{\hat{r}}_{(0)}\psi^{I}_{(0)i+}\ ,\nn
 \\
\delta\psi^{I}_{(0)i+}&=&
\ft12{\cal K}^{ab}_{(0)i}\gamma_{ab}\epsilon^{I}_+
                  +\sqrt{2}A^{~IJ}_{(0)i}\epsilon^{J}_+\ ,
\eea
we then deduce that the vanishing of $\psi^{I}_{(0)i+}$
implies $A^{~IJ}_{(0)i}=0$. The supersymmetry variation of
$A^{~IJ}_{(0)i}$, given by
\bea
\delta A^{IJ}_{(0)i}&=&-\Big(\cos2\omega\bar{\epsilon}^{K}_+
 \gamma_{(0)i}\chi^{IJK}_{+}+{\rm i}\sin2\omega\bar{\epsilon}^{K}_+
\gamma_{(0)i}\gamma_5\chi^{IJK}_{-}\Big)+\cdots\ ,
\nn\\
\delta A^{IJ}_{(1)i}&=&-{\cal S}_{(1)}^{IJKL}\bar{\epsilon}^{M}_+\gamma_{(0)i}\chi^{KLM}_{+}
-\im{\cal P}_{(1)}^{IJKL}\bar{\epsilon}^{M}_+\gamma_{(0)i}\gamma_5\chi^{KLM}_{-}
-2\sqrt{2}\bar{\epsilon}^{[I}_+\psi^{J]}_{i(3)-}
\nn\\
&&+D_{i}(A_{(0)})\Big(\cos2\omega\bar{\epsilon}_{K+}\chi^{IJK}_{-}
+{\rm i}\sin2\omega\bar{\epsilon}_{K+}\gamma_5\chi^{IJK}_+\Big)+\cdots\ ,
\eea
will then require\footnote{The ellipses refer to term depending on
${\psi}^{r}_{(0)i+}$ which vanish for the Dirichlet boundary conditions
that we impose on the gravitini. }
\be
\cos2\omega\, \chi^{IJK}_{+}+{\rm i}\sin2\omega\, \gamma_5\chi^{IJK}_{-}=0\ .
\label{bcdchiIJK}
\ee
The variation of the above equation should also vanish. Using
\bea
\delta\chi^{IJK}_{+}&=&-{\cal S}^{IJKL}_{(2)}\epsilon^L_++2\im  {\cal P}^{IJKL}_{(1)}\gamma_5\epsilon^L_--\im\slashed{D}{\cal P}_{(1)}^{IJKL}\gamma_5\epsilon^L_+\ ,
\nn\\
\delta\chi^{IJK}_{-}&=&2{{\cal S}}^{IJKL}_{(1)}\epsilon^L_--{\rm i}{\cal P}^{IJKL}_{(2)}\gamma_5\epsilon^L_++\slashed{D}{{\cal S}}_{(1)}^{IJKL}\epsilon^L_+\ ,
\eea
we deduce that
\bea
\cos2\omega{\cal P}^{IJKL}_{(1)}+\sin2\omega{{\cal S}}^{IJKL}_{(1)}&=&0\Rightarrow
-\cos2\omega{\cal P}^{IJ}_{(1)}+\sin2\omega{{\cal S}}^{IJ}_{(1)}=0\ ,
\nn\\
\cos2\omega{{\cal S}}^{IJKL}_{(2)}-\sin2\omega{\cal P}^{IJKL}_{(2)}&=&0
 \Rightarrow \cos2\omega{{\cal S}}^{IJ}_{(2)}+\sin2\omega{\cal P}^{IJ}_{(2)}=0\ ,
\label{bcdscalar}
\eea
which will further imply
\be
\cos2\omega{\chi}^{I}_{+}-{\rm i}\sin2\omega\gamma_5{\chi}^{I}_{-}=0\ ,
\label{bcdchiI}
\ee
according to the variation of ${{\cal S}}^{IJ}$ and ${\cal P}^{IJ}$
\bea
\delta{{\cal S}}^{IJ}_{(1)}&=&2\bar{\epsilon}^{[I}_+{\chi}_{-}^{J]}+\ft1{3!}\varepsilon^{IJKLMN}\bar{{\epsilon}}^{K}_+{\chi}_{-}^{LMN}\ ,
\nn\\
\delta{\cal P}^{IJ}_{(1)}&=&-2\im\bar{{\epsilon}}^{[I}_+\gamma_5{\chi}_{+}^{J]}
+\ft{\im}{3!}\varepsilon^{IJKLMN}\bar{\epsilon}^{K}_+\gamma_5{\chi}_{+}^{LMN}\ ,
\nn\\
\delta{{\cal S}}^{IJ}_{(2)}&=&2\Big(\bar{\epsilon}^{[I}_-{\chi}_{+}^{J]}+\bar{\epsilon}^{[I}_+\slashed{D}{\chi}_{+}^{J]}\Big)
+\ft1{3!}\varepsilon^{IJKLMN}\Big(\bar{\epsilon}^{K}_-{\chi}_{+}^{LMN}+\bar{\epsilon}^{K}_+\slashed{D}{\chi}_{+}^{LMN}\Big)\ ,
\nn\\
\delta{\cal P}^{IJ}_{(2)}&=&
-2\im\Big(\bar{\epsilon}^{[I}_-\gamma_5{\chi}_{-}^{J]}-\bar{\epsilon}^{[I}_+\gamma_5\slashed{D}{\chi}_{-}^{J]}\Big)
+\ft{\im}{3!}\varepsilon^{IJKLMN}\Big(\bar{\epsilon}^{K}_-\gamma_5{\chi}_{-}^{LMN}
-\bar{\epsilon}^{K}_+\gamma_5\slashed{D}{\chi}_{-}^{LMN}\Big)\ .
\eea
It can be checked that (\ref{bcdscalar}) and (\ref{bcdchiI}) are consistent with the variation of $\chi^I$
\bea
\delta{\chi}^{I}_{+}&=&-{\cal S}^{IJ}_{(2)}{\epsilon}^J_++2\im{\cal P}^{IJ}_{(1)}\gamma_5{\epsilon}^J_--\im\slashed{D}{\cal P}_{(1)}^{IJ}\gamma_5{\epsilon}^J_+,\nn\\
\delta{\chi}^{I}_{-}&=&2{{\cal S}}^{IJ}_{(1)}{\epsilon}^J_--{\rm i}{\cal P}^{IJ}_{(2)}\gamma_5{\epsilon}^J_+
+\slashed{D}{{\cal S}}_{(1)}^{IJ}{\epsilon}^J_+\ .
\eea
Finally, given the boundary conditions for the spin-$\ft32$, spin-$\ft12$ (\ref{bcdchiIJK}), (\ref{bcdchiI}) and
spin-0 fields (\ref{bcdscalar}), we can derive the admissible boundary
condition for the $U(1)$ gauge field that preserve ${\cal N}=6$ supersymmetry. Using
\bea
\delta A_{(0)i}&=&
-\Big(\cos2\omega\bar{\epsilon}^{I}_+\gamma_{(0)i}{\chi}^{I}_{+}
+{\rm i}\sin2\omega\bar{\epsilon}^{I}_+\gamma_{(0)i}\gamma_5{\chi}^{I}_{-}\Big)+\cdots\ ,
\nn\\
\delta A_{(1)i}&=&-{{\cal S}}_{(1)}^{IJ}\bar{\epsilon}^{K}_+\gamma_{(0)i}\widetilde{\chi}^{IJK}_{+}
-\im{\cal P}_{(1)}^{IJ}\bar{\epsilon}^{K}_+\gamma_{(0)i}\gamma_5{\chi}^{IJK}_{-}
\nn\\
&&+\partial_i\Big(\cos2\omega\bar{\epsilon}_{I+}{\chi}^{I}_{-}+{\rm i}\sin2\omega\bar{\epsilon}_{I+}\gamma_5{\chi}^{I}_+\Big)+\cdots\ ,
\eea
one can see that when $\omega\neq0$, the only boundary condition
preserving the ${\cal N}=6$ supersymmetry is given by
\be
A_{(1)\mu}=0,\qquad \omega=\frac{\pi}{8}\ .
\ee
In other words, the only case within the class of $\omega$-deformed
${\cal N}=6$ supergravities where there can exist consistent boundary
conditions that preserve the ${\cal N}=6$ supersymmetry is when
$\omega=\pi/8$.  The $U(1)$ gauge field must then satisfy the Neumann,
rather than Dirichlet, boundary condition.

If we redefine the complex scalar
\be
\widetilde{\varphi}^{IJ}\equiv e^{-2\im \omega}\varphi^{IJ}\ ,
\label{redefvarphi}
\ee
the ${\cal N}=6$ boundary condition can be summarized as
\bea
&&e^{\hat{r}}_{i}=\delta^{\hat{r}}_{i},~{\psi}^{I}_{(0)i}=0\ ,
~A^{IJ}_{(0)i}=0,~A_{(1)i}=0,~\widetilde{{\cal S}}^{IJ}_{(1)}=0\ ,
~\widetilde{\cal P}^{IJ}_{(2)}=0\ ,
\nn\\
&&\cos2\omega\chi^{IJK}_{+}+{\rm i}\sin2\omega\gamma_5\chi^{IJK}_{-}=0\ ,
\quad \cos2\omega{\chi}^{I}_{+}-{\rm i}\sin2\omega\gamma_5{\chi}^{I}_{-}=0\ ,
\label{N=6bndcon}
\eea
with the condition also that $\omega=\pi/8$.

\section{$\omega$-Dependent 3-Point Tree Graphs}

   One of our primary goals in this paper is to identify and compute
the simplest correlation functions that are sensitive to the
$\omega$-deformations of the bulk supergravity theory, in order to
gain insights into the effects of the deformations on the boundary conformal
field theory.  Our starting point is to study the expansion of the
four-dimensional fields around the trivial AdS$_4$ vacuum of the
${\cal N}=8$ $\omega$-deformed gauged supergravity.  For the reasons discussed
already, our focus will be on the consistent truncation of the
${\cal N}=8$ theory to ${\cal N}=6$, for which, as we showed in section 4,
we can impose supersymmetric boundary conditions in which the graviton and
gravitini obey standard Dirichlet asymptotics.   The scalar potential
of the ${\cal N}=6$ truncation is independent of the deformation parameter
$\omega$, and so within the bosonic sector this leaves the coupling of the
scalars in the gauge-field kinetic terms as the remaining place where
$\omega$-dependence can enter.  In appendix A, we present the expansion
of the the gauge-field kinetic terms of the $\omega$-deformed
${\cal N}=8$ theory up to quartic order in the bosonic fields.  From this
expansion, it is evident that there is in fact $\omega$ dependence in the
trilinear couplings of two gauge fields with a scalar field, and thus
for our present purposes it suffices to focus on these terms of cubic order
in the bosonic fields.

   The $\omega$ dependence in
the $\varepsilon_{IJKLMN}\varphi^{IJ}F^{KL}F^{MN}$ terms can be absorbed
into a redefinition of $\varphi^{IJ}$, as in (\ref{redefvarphi}).
A further advantage of
using the redefined $\widetilde\varphi^{IJ}$  is that the real and
imaginary parts of $\widetilde{\varphi}^{IJ}$ then satisfy definite boundary
conditions (\ref{N=6bndcon}), whereas the boundary conditions imposed on
the original scalars $\varphi^{IJ}$ involve $\omega$-dependent linear
combinations of the real and imaginary parts of $\varphi^{IJ}$.

   After taking the redefinition into account, we are left
with two $\omega$-dependent vertices at the trilinear order.  After
truncating to the fields of ${\cal N}=6$ supergravity, as discussed in
section 3, these are given by
\bea
&&1)\qquad -\sqrt{2}\int \frac{d^4z}{z_0^4}{\rm Re}
(e^{4{\rm i}\omega} \, {\widetilde\varphi}^{IJ})
\partial_{[\mu} A_{\nu]}{}^{IJ}
\, \partial_{\mu} A_{\nu }\ ,
\nn\\
&&2)\qquad -\frac{\im}{\sqrt{2}}\int \frac{d^4z}{z_0^4} {\rm Im}
(e^{4 i \omega}\, \widetilde{\varphi}^{IJ})  \, \epsilon_{\mu \nu \rho \sigma}
\partial_{\mu} A_{\nu}{}^{IJ}\partial_{\rho} A_{\sigma}\ .
\label{2amps}
\eea
In the above expressions, we do not distinguish the lower and upper indices since in this section we work with Euclidean signature. As we discussed in the previous section, for non-zero values of the
$\omega$ parameter ${\cal N}=6$ supersymmetry
of the boundary conditions
requires $\omega=\pi/8$, and specific asymptotic falloff behaviors for
the scalars near the boundary of AdS. Explicitly,
$\widetilde{{\cS}}^{IJ}(z)\sim z_0^2 \widetilde{{\cS}}^{IJ}(\vec{z}) $
and $\widetilde{{\cP}}^{IJ}(z)\sim z_0 \widetilde{{\cP}}^{IJ}(\vec{z})$,
which means $\Delta=2$ in the bulk to boundary propagator of
$\widetilde{{\cS}}^{IJ}$ and $\Delta=1$ in the bulk to boundary propagator
of $\widetilde{{\cP}}^{IJ}$.  We shall, however, keep our discussion more
general, and leave the conformal dimension $\Delta$ for the
scalars $\widetilde{{\cS}}^{IJ}$ and $\widetilde{{\cP}}^{IJ}$ arbitrary for
now.

According to the AdS/CFT dictionary, the cubic interactions in the
supergravity Lagrangian are associated with certain 3-point correlation
functions in the boundary CFT. The mapping from the bulk interaction to
the correlators on the boundary is represented by the Witten diagram:
\begin{center}
\begin{align*}
\begin{xy}
(50,20)*\xycircle<40pt>{},
(50,20)*{};
(38,27.3)*=0{+} *+!DR=0{A_{i}^{KL}(\vec x_2)} **@{~},
(62,27.3)*=0{+} *+!DL=0{A_{j}(\vec x_3)} **@{~},
(50,6)*=0{\times} *++!U=0{\widetilde{\varphi}^{IJ}(\vec x_1)} **@{--},
\end{xy}
\end{align*}
\\[5mm]
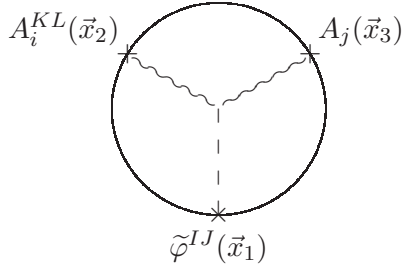
{\it Fig.~1. Witten diagram corresponding to the $\omega$-dependent 
bulk 3-point interactions.}
\end{center}

The 3-point amplitude corresponding to the first bulk cubic interaction
in (\ref{2amps}) can be expressed as
\bea
T^{(1)IJ,KL}_{ij}(\vec x_1,\vec x_2,\vec x_3)&=&-\frac{1}{\sqrt{2}}
(\delta^{IK}\delta^{JL}-\delta^{IL}\delta^{JK})
\nn\\
&&\times  \int \frac{dz_0d^3\vec z}{z_0^4}K_{\Delta}( z,\vec x_1)
g^{\mu\rho}g^{\nu\sigma}\partial_{[\mu}G_{\nu] i}(z,\vec x_2)
\partial_{\rho}\widetilde{G}_{\sigma j}(z,\vec x_3)\ ,
\label{aaphi1}
\eea
in which the $K_{\Delta}(z,\vec x)$ is the bulk to boundary 
propagator associated with the scalar field of conformal dimension $\Delta$,
\be
K_{\Delta}(z,\vec x)=c_{\Delta}\Big(\frac{z_0}{z_0^2+(\vec z-\vec x)^2}
\Big)^{\Delta}\ ,
\quad \quad c_{\Delta}=\frac{\Gamma(\Delta)}{\pi^{\frac{3}{2}}
\Gamma(\Delta-\frac{3}{2})}\ .
\ee
$G_{\mu i}(z,\vec x)$ is the bulk to boundary propagator associated
with $A_{i}^{IJ}$. Since $A_{i}^{IJ}$ satisfies Dirichlet boundary
conditions, its bulk to boundary propagator takes the standard
form \cite{Freedman:1998tz}
\be
G_{\mu i}(z,\vec x)=c_3\frac{z_0}{[z_0^2+(\vec z-\vec x)^2]^2}
J_{\mu i}(z-\vec x),\quad c_3=\frac{2}{\pi^2}\ ,
\ee
where
\be
J_{\mu\nu}(x-y)=\delta_{\mu\nu} - \fft{2(x-y)_\mu\, (x-y)_\nu}{|x-y|^2}\ .
\ee
It follows that  $G_{\mu i}(z,\vec x)$ satisfies
\be
\quad \partial_{[\mu}G_{\nu]i}(z,\vec x)=
\frac{2}{\pi^2[z_0^2+(\vec z-\vec x)^2]^2}\, J_{0[\mu}(z-\vec x)
J_{\nu]i}(z-\vec x)\ .
\ee
It can also be checked that
\be
\frac{\partial}{\partial x_i}\partial_{[\mu}G_{\nu]i}(z,\vec x)=0\ ,
\ee
which implies that $A_{i}^{IJ}$ is dual to a conserved current of
dimension 2. $\widetilde{G}_{\mu i}(z,\vec x)$ is the bulk to boundary propagator
associated with $A_{i}$.

Since the bulk $U(1)$ gauge field $A_{i}$
satisfies instead a Neumann boundary condition, its bulk to boundary
propagator takes a different form \cite{D'Hoker:1999jc}.
 It is shown in \cite{D'Hoker:1999jc} that up to a pure gauge term, the
propagator for a $U(1)$ gauge field in AdS$_{d+1}$  can be expressed as
\be
G_{\mu\nu}(z,w)=-F(u)\, \fft{\partial^2 u}{\partial z^\mu\partial w^\nu}
 \,,
\qquad u\equiv \frac{(z-w)^2}{2z_0w_0}\ ,
\ee
where $F(u)$ satisfies
\be
u(u+2)F''+(d+1)(1+u)F'+(d-1)F=0\ .
\ee
Up to a proportionality constant, the two independent solutions of this
equation are given by
\be
F_1(u)=\frac1{u(u+2)}\,,\qquad F_2(u)=\frac1u\ .
\ee
The propagator associated with the usual Dirichlet boundary
condition can be obtained by choosing
$F(u)\propto F_1$. If instead $F(u)$ is
chosen to be proportional to $F_2(u)$, we obtain the $U(1)$
gauge boson propagator
associated with the Neumann boundary condition, which we shall
 denote by $\widetilde{G}_{\mu\nu}$.  It is given by
\be
\widetilde{G}_{\mu i}(z,\vec x)=
\tilde{c}_3\frac{1}{[ z_0^2+(\vec z-\vec x)^2]}(\delta_{\mu i}+
\frac{(\vec x-\vec z)_i}{z_0}\delta_{\mu 0}),\quad
\tilde{c}_3=-\frac{1}{2\pi^2}\ ,
\ee
with its curl given by
\be
\partial_{[\mu}\widetilde{G}_{\nu]i}(z,\vec x)=
-\frac{1}{2\pi^2z_0[ z_0^2+(\vec z-\vec x)^2]}J_{0[\mu}(z-\vec x)
J_{\nu]i}(z-\vec x)\ .
\ee
The constant
$\tilde{c}_3$ is fixed by requiring that
$\widetilde{G}_{\mu i}(z,\vec x)\rightarrow
z_0\delta_{ij}\delta^3(\vec z-\vec x)$ when $z_0\rightarrow 0$.

    It should be noted that unlike the Dirichlet propagator
$G_{\mu i}(z,\vec x)$, the Neumann propagator has
\be
\frac{\partial}{\partial x_i}\partial_{[\mu}\widetilde{G}_{\nu]i}(z,\vec x)
\neq0\ .
\ee
This means that the holographic dual of $A_i$ cannot be a conserved current.
On the other hand, the 2-point amplitude associated with two $A_i$ fields
reveals that
the dual of $A_i$ should have dimension 1, which is below the unitary
bound for a spin-1 operator in ${\rm CFT}_3$. However, 1 is the correct
dimension for
a Chern-Simons gauge field. Following \cite{Marolf:2006nd}, these facts
lead to the conclusion that the holographic dual of $A_i$ is a
dynamical $U(1)$ Chern-Simons gauge field.

Because of the translation invariance in the 3-dimensional boundary
directions, the 3-point amplitude derived from the first cubic vertex
depends only on the difference of boundary coordinates. Thus
\be
T^{(1)IJ,KL}_{ij}(\vec x_1,\vec x_2,\vec x_3)=T^{(1)IJ,KL}_{ij}(\vec x_{13},
\vec x_{23},0)\ ,
\ee
where $\vec x_{13}=\vec x_1-\vec x_3$ and $\vec x_{23}=\vec x_2-\vec x_3$.
For simplicity, we shall first compute the amplitude
\be
T^{(1)IJ,KL}_{ij}(\vec w,\vec x,0)\ ,
\label{amp1}
\ee
and later replace $\vec w$ by $\vec x_{13}$ and $\vec x$ by
$\vec x_{23}$. To compute (\ref{amp1}), we follow the strategy of
\cite{Freedman:1998tz} by expressing (\ref{amp1}) as
\be
T^{(1)IJ,KL}_{ij}(\vec w,\vec x,\vec y)|_{|y|\rightarrow0}\ .
\ee
Then we use the inversion trick,
\be
z_{\mu}=\frac{z'_{\mu}}{z'^2},\quad \vec w=\frac{\vec w'}{|w'|^2},
\quad \vec x=\frac{\vec x'}{|x'|^2},\quad \vec y=\frac{\vec y'}{|y'|^2}\ .
\ee
Under the inversion of coordinates, the propagators transform covariantly
as
\bea
K_{\Delta}(z,\vec w)&=&K_{\Delta}( z',\vec w')|w'|^{2\Delta}\ ,
\nn\\
\partial_{[\mu}G_{\nu]i}(z,\vec x)&=&(z')^2J_{\mu\rho}(z')\cdot (z')^2
J_{\nu\sigma}(z')\cdot(\vec x')^4J_{ki}(\vec x)\cdot
\partial'_{[\rho}G_{\sigma]k}(z',\vec x')\ ,
\nn\\
\partial_{[\mu}\widetilde{G}_{\nu]i}(z,\vec x)&=
&(z')^2J_{\mu\rho}(z')\cdot (z')^2J_{\nu\sigma}(z')\cdot(\vec x')^2
J_{ki}(\vec x)\cdot \partial'_{[\rho}\widetilde{G}_{\sigma]k}(z',\vec x')\ .
\eea
After some algebra, the scalar integral in (\ref{aaphi1}) can be
simplified to give
\bea
&& c_{\Delta}\tilde{c}_3\int dz'_0d^3\vec z'
\frac{z_0^{'\Delta}}{[z^{'2}_0+(\vec z'-\vec w')^2]^{\Delta}}(\vec x')^4
J_{ki}(\vec x')\partial'_{[0}G_{j]k}(z',\vec x')
\frac1{z_0'}(\vec w')^{2\Delta}
\nn\\
&=&-\frac{\Gamma(\Delta)}{\pi^{\frac{11}2}
\Gamma(\Delta-\frac32)}(\vec x')^4(\vec w')^{2\Delta}J_{ki}(\vec x')
\nn\\
&&\qquad \times \int \frac{dz'_0d^3\vec z'}{z'_0}\Big(
\frac{z'_0}{(z'-\vec w')^2}\Big)^{\Delta}\frac{\partial}{\partial z'_{[0}}
\Big(\frac{z'_0}{(z'-\vec x')^2}\Big)\frac{\partial}{\partial z'_{j]}}
\Big(\frac{(z'-\vec x')_k}{(z'-\vec x')^2}\Big)\ .
\eea
The above expression can be computed by using two integral formulae.
The first is given in \cite{Freedman:1998tz}, namely
\bea
&&\int^{\infty}_0 dz_0\int d^3\vec z\frac{z_0^a}{[z_0^2+
   (\vec z-\vec x)^2]^b[z_0^2+(\vec z-\vec y)^2]^c}\equiv
 I[a,b,c,3]|\vec x-\vec y|^{4+a-2b-2c} ,
 \nn\\
&&I[a,b,c,3]=\frac{\pi^{\frac32}}{2}\frac{\Gamma[\frac{a+1}2]
  \Gamma[b+c-\frac{a}2-2]\Gamma[2+\frac{a}2-b]\Gamma[2+\frac{a}2-c]}{
   \Gamma[b]\Gamma[c]\Gamma[4+ a-b-c]}\ .
\label{int1}
\eea
The second takes the form
\bea
&&\int_0^{\infty} dz_0\int d^3\vec z\frac{z_0^a(\vec z-\vec y)_i}{[z_0^2+
  (\vec z-\vec x)^2]^b[z_0^2+(\vec z-\vec y)^2]^c}=
\widetilde{I}[a,b,c,3]\frac{(\vec x-\vec y)_i}{|\vec x-\vec y|^{2b+2c-a-4}}\ ,
\nn\\
&&\widetilde{I}[a,b,c,3]=\frac{\pi^{\frac32}}{2}\frac{\Gamma[\frac{a +1}2]
\Gamma[3+\frac{a}2-c]\Gamma[2+\frac{a}2-b]\Gamma[b+c-\frac{a}2-2]}{
\Gamma[b]\Gamma[c]\Gamma[5+a-b-c]}\ .
\label{int2}
\eea
In fact for $c>2$ the second integral can be expressed as the derivative
of the first with respect to $y_i$. For generic values of $c$,
integral (\ref{int2}) can be directly computed by using Feynman integral
techniques with two denominators. After performing the $z$ integral,
we obtain
\bea
T^{(1)IJ,KL}_{ij}(\vec w,\vec x,0)&=&\frac{\Gamma[\frac{1+\Delta}2]^2
   \Gamma[\frac{\Delta}2]\Gamma[\frac{3-\Delta}2]}{16\sqrt{2}\pi^4
 \Gamma[\Delta-\frac32]}(\delta^{IK}\delta^{JL}-\delta^{IL}\delta^{JK})
 \nn\\
&&\times\Big((\Delta-3)J_{ij}(\vec x')+(\Delta+1)J_{ik}(\vec x')
J_{kj}(\vec w'-\vec x')\Big)\frac{|\vec x'|^4|\vec w'|^{2\Delta}}{
 |\vec w'-\vec x'|^{\Delta+1}}
 \nn\\
&=&\frac{\Gamma[\frac{1+\Delta}2]^2\Gamma[\frac{\Delta}2]\Gamma[
\frac{3-\Delta}2]}{16\sqrt{2}\pi^4\Gamma[\Delta-\frac32]}
(\delta^{IK}\delta^{JL}-\delta^{IL}\delta^{JK})
\nn\\
&&\times\frac{(\Delta-3)J_{ij}(\vec x)+(\Delta+1)
 J_{ik}(\vec x-\vec w)J_{kj}(\vec w)}{|\vec x|^{3-\Delta}|\vec w|^{\Delta-1}|
\vec w-\vec x|^{\Delta+1}}\ .
\eea
It can be checked that
\be
\frac{\partial}{\partial x_j}T^{(1)IJ,KL}_{ij}(\vec w,\vec x,0)=0\ .
\ee
In terms of the original $\vec x_i$ coordinates,
\bea
T^{(1)IJ,KL}_{ij}(\vec x_1,\vec x_2,\vec x_3)&=&
 \frac{\Gamma[\frac{1+\Delta}2]^2\Gamma[\frac{\Delta}2]
 \Gamma[\frac{3-\Delta}2]}{16\sqrt{2}\pi^4\Gamma[\Delta-\frac32]}
(\delta^{IK}\delta^{JL}-\delta^{IL}\delta^{JK})
\nn\\
&&\!\!\!\!\!\!\!\! \!\!\!\!\!
\times\frac{(\Delta-3)J_{ij}(\vec x_2-\vec x_3)+
(\Delta+1)J_{ik}(\vec x_2-\vec x_1)J_{kj}(\vec x_1-\vec x_3)}{
|\vec x_2-\vec x_3|^{3-\Delta}|\vec x_1-\vec x_3|^{\Delta-1}
  |\vec x_1-\vec x_2|^{\Delta+1}}\ .
\eea

The second $\omega$-dependent cubic vertex in (\ref{2amps})
leads to another 3-point boundary amplitude,
\bea
T^{(2)IJ,KL}_{ij}(\vec x_1,\vec x_2,\vec x_3)&=&-
\frac{\im}{2\sqrt{2}}(\delta^{IK}\delta^{JL}-\delta^{IL}\delta^{JK})
\nn\\
&&\times  \int \frac{dz_0d^3\vec z}{z_0^4}K_{\Delta}( z,\vec x_1)
\epsilon^{\mu\nu\rho\sigma}\partial_{\mu}G_{\nu i}(z,\vec x_2)
\partial_{\rho}\widetilde{G}_{\sigma j}(z,\vec x_3)\ .
\label{aaphi2}
\eea
Following the same strategy, we shall compute
\be
T^{(2)IJ,KL}_{ij}(\vec w,\vec x,0)\ ,
\label{amp2}
\ee
and replace $w$ by $\vec x_{13}$ and $x$ by $\vec x_{23}$ at the final stage
of the calculation. Utilizing again the inversion trick and the properties
of the various propagators under the inversion of coordinates,
the scalar integral in the amplitude (\ref{aaphi2}) can be expressed as
\bea
&&c_{\Delta}\tilde{c}_3\int \frac{dz'_0d^3\vec z'}{z_0^{'4}}
\frac{z_0^{'\Delta}}{[z^{'2}_0+(\vec z'-\vec w')^2]^{\Delta}}
\epsilon^{mnj0}(\vec x')^4J_{ki}(\vec x')\partial_m'G_{nk}(z',\vec x')
\frac1{z_0'}(\vec w')^{2\Delta}
\nn\\
&=&\frac{2\Gamma(\Delta)\varepsilon^{mkj0} }{\pi^{\frac{11}2}
   \Gamma(\Delta-\frac32)}
\int dz'_0d^3\vec z'
\frac{z_0^{'\Delta}}{[z^{'2}_0+(\vec z'-\vec w')^2]^{\Delta}}
\frac{(\vec x')^4J_{ki}(\vec x')
  (\vec z'-\vec x')_m(\vec w')^{2\Delta}}{[z^{'2}_0+(\vec z'-\vec x')^2]^3}\ .
\eea
In the derivation of the above results, the following equalities have been used
\bea
&&\epsilon^{\mu\nu\rho\sigma}J_{\mu\lambda}(z')J_{\nu\tau}(z')
J_{\rho\gamma}(z')J_{\sigma\eta}(z')=-\epsilon^{\lambda\tau\gamma\eta}\ ,
\nn\\
&&J_{lj}(\vec y')J_{\eta l}(\vec z'-\vec y')J_{0\gamma}(\vec z'-\vec y')
 |_{|y'|\rightarrow\infty}=\delta_{j\eta}\delta_{0\gamma}\,.
\eea
Using the previous integral formulae, we obtain
\bea
T^{(2)IJ,KL}_{ij}(\vec w,\vec x,0)&=&-\im\frac{\Gamma[\frac{1+\Delta}2]\Gamma
[\frac{\Delta}2]^2\Gamma[2-\frac{\Delta}2]}{8\sqrt{2}\pi^4
\Gamma[\Delta-\frac32]}(\delta^{IK}\delta^{JL}-\delta^{IL}\delta^{JK}) 
\nn\\
&&\times (\vec x')^4(\vec w')^{2\Delta}\varepsilon_{mkj}
J_{ki}(\vec x')\partial_{x'_{m}}|\vec w'-\vec x'|^{-\Delta}
\nn\\
&=&-\im \frac{\Delta\Gamma[\frac{1+\Delta}2]\Gamma
[\frac{\Delta}2]^2\Gamma[2-\frac{\Delta}2]}{8\sqrt{2}\pi^4
\Gamma[\Delta-\frac32]}(\delta^{IK}\delta^{JL}-\delta^{IL}\delta^{JK})
\nn\\
&&\times \frac{|\vec w|^{2-\Delta}}{
  |\vec w-\vec x|^{\Delta}|\vec x|^{4-\Delta}}\varepsilon_{mkj}
J_{ki}(\vec x)J_{m\ell}(\vec w)\Big[\frac{(w-x)_{\ell}}{
         |\vec w-\vec x|^2}-\frac{w_{\ell}}{|\vec w|^2}\Big]\ .
\eea
In terms of the original $x_i$ coordinates, the second 3-point amplitude is given by
\bea
&&T^{(2)IJ,KL}_{ij}(\vec x_1,\vec x_2,\vec x_3)=
\im \frac{\Gamma[\frac{1+\Delta}2]\Gamma
[\frac{\Delta}2]\Gamma[\frac{\Delta}2+1]\Gamma[2-\frac{\Delta}2]}{4\sqrt{2}\pi^4
\Gamma[\Delta-\frac32]}(\delta^{IK}\delta^{JL}-\delta^{IL}\delta^{JK})
\nn\\
&&\quad \times \frac{|\vec x_1-\vec x_3|^{2-\Delta} \varepsilon_{mkj}}{
 |\vec x_1-\vec x_2|^{\Delta+2}
|\vec x_2-\vec x_3|^{2-\Delta}} J_{ki}(\vec x_2-\vec x_3)
\Big[\frac{(x_3-x_1)_{m}}{
|\vec x_3-\vec x_1|^2}-\frac{(x_3- x_2)_{m}}{|\vec x_3-\vec x_2|^2}
\Big]\ .
\eea
To arrive at the result above, we have used a useful formula given below
\bea
J_{m\ell}(\vec x_3-\vec x_1)\Big[\frac{(x_1-x_2)_{\ell}}{
|\vec x_1-\vec x_2|^2}-\frac{(x_1- x_3)_{\ell}}{|\vec x_1-\vec x_3|^2}\Big]=-\frac{|\vec x_3-\vec x_2|^2}{|\vec x_1-\vec x_2|^2}
\Big[\frac{(x_3-x_1)_{m}}{|\vec x_3-\vec x_1|^2}-\frac{(x_3- x_2)_{m}}{|\vec x_3-\vec x_2|^2}\Big]\ .
\eea
%

\section{Interpretation in the Dual Theory}

In this section we shall discuss how our results for amplitudes
in the $\omega$-deformed
bulk theory at $\omega=\pi/8$ are related to a certain operation on the
$U(1)\times U(1)$ sector of the $U(N)_k\times U(N)_{-k}$ ABJM theory.
In the bulk, we shall denote the $U(1)$ gauge field in the $\omega=0$
theory by $A'_{\mu}$, to distinguish it from $A_{\mu}$ in the
deformed theory.

The holographic 2-point function associated the $U(1)$ bulk gauge field
for the $\omega=0$ case is given by \cite{Freedman:1998tz},
\be
\frac{\delta^2 S[A']}{\delta A'_{(0)i}(\vec x_1)\ \delta A'_{(0)j}(\vec x_2)}\Bigg|_{\omega=0}=\langle J_i(\vec x_1)J_j(\vec x_2)\rangle\Big|_{\omega=0}=\frac1{\pi^2}\frac{J_{ij}}{|\vec x_1-\vec x_2|^4}\  .
\ee
In the $\omega=\pi/8$ theory, recalling that the bulk $U(1)$ gauge field
obeys a Neumann boundary condition, we have
\be
\frac{\delta^2 S[A]}{\delta A_{(1)i}(\vec x_1)\ \delta A_{(1)j}(\vec x_2)}\Bigg|_{\omega=\pi/8}=\langle A_{(0)i}(\vec x_1)A_{(0)j}(\vec x_2)\rangle\Big|_{\omega=\pi/8}=\frac1{4\pi^2}\frac{\delta_{ij}}{|\vec x_1-\vec x_2|^2}\  .
\ee
It is easy to check that 
\be
\langle J^{\rm top}_i(\vec x_1)J^{\rm top}_j(\vec x_2)\rangle\Big|_{\omega=\pi/8}=\langle J_i(\vec x_1)J_j(\vec x_2)\rangle\Big|_{\omega=0}\ ,
\label{2J}
\ee
where we have defined the topological current to be
\be
J^{\rm top}_i=\im \varepsilon_{ijk}\partial_j A_{(0)k}\ .
\ee
This relation can be understood as the electric-magnetic rotation in the
bulk theory as follows. From (\ref{dualU1}), it can be seen that
the $U(1)$ gauge fields in $\omega=0$ and $\omega=\pi/8$ theories
are related on-shell by
\be
F'_{\mu\nu}=-G_{\mu\nu}\ ,
\label{6.5}
\ee
where, in Minkowski signature, we have
\be
G_{\mu\nu}=\ft12\epsilon_{\mu\nu\alpha\beta}F^{\alpha\beta}+\sqrt{2}\,
{\rm Re}(e^{2\im \omega}\phi^{IJ})\epsilon_{\mu\nu\alpha\beta}\partial^{\alpha}A^{\beta}
+\sqrt{2}\, {\rm Im} (e^{2\im \omega}\phi^{IJ})\partial_{\mu}A_{\nu}+\cdots\ .
\ee
Substituting the FG expansions (\ref{FGexp0}), applicable for both
the $A_i$ and $A_i'$ fields, into (\ref{6.5}), we obtain

\be
A'_{(1)i}=\varepsilon_i^{~jk}\partial_j A_{(0)k},
 \quad A_{(1)i}=-\varepsilon_i^{~jk}\partial_j A'_{(0)k}\ .
\label{5.47}
\ee
In the Euclidean signature, the first equation  implies
\be
\frac{\delta S[A']}{\delta A'_{(0)i}}\Bigg|_{\omega=0}=
\im \varepsilon_{ijk}\partial_j \frac{\delta S[A]}{
   \delta A_{(1)k}}\Bigg|_{\omega=\pi/8}\ .
\label{varyaction}
\ee
In the $\omega=0$ case, $A'_{(0)i}$ is treated as the source and
$A'_{(1)i}$ is the VEV, whereas in the $\omega=\pi/8$ case, the roles
of $A_{(0)i}$ and $A_{(1)i}$ interchange, with $A_{(0)i}$ becoming
the VEV and $A_{(1)i}$ playing the role of the source.
Differentiating (\ref{varyaction}) with respect to $A'_{(0)j}$ and
using (\ref{5.47}), we derive the relation (\ref{2J}). In fact, it
follows from (\ref{varyaction}) that $(n+m)$-point
functions obey, schematically,
\be
\langle \cO_1\cdots\cO_nJ^{\rm top}_1\cdots J^{\rm top}_n\rangle\Big|_{
\omega=\pi/8}=\langle \cO_1\cdots\cO_nJ_1\cdots J_n
  \rangle\Big|_{\omega=0}\ ,
  \label{OOJJ}
  \ee
where the $\cO$ denote scalar, vector or tensor primary operators in the 
ABJM model.

     As a non-trivial check of this relation, we now examine the
3-point correlation functions calculated in the previous section.
In the $\omega=\pi/8$ deformed theory, we found
\bea
&&\langle {\cO}^{\Delta=1}(\vec x_1)^{IJ}J_i(\vec x_2)^{KL}A_j(\vec x_3)
\rangle\Big|_{\omega=\pi/8}=-T^{(1)IJ,KL}_{ij,\Delta=1}\ ,
\nn\\
&&\langle {\cO}^{\Delta=2}(\vec x_1)^{IJ}J_i(\vec x_2)^{KL}A_j(\vec x_3)
\rangle\Big|_{\omega=\pi/8}=T^{(2)IJ,KL}_{ij,\Delta=2}\ .
\label{5.33}
\eea
In the $\omega=0$ theory, the cubic-interaction vertices
(\ref{2amps}) take the form
\bea
&&1)\qquad -\sqrt{2}\int \frac{d^4z}{z_0^4}{\cS}^{IJ}
\partial_{[\mu} A_{\nu]}{}^{IJ}
\, \partial_{\mu} A_{\nu }\ ,
\nn\\
&&2)\qquad -\frac{\im}{\sqrt{2}}\int \frac{d^4z}{z_0^4} {\cP}^{IJ}  \, \epsilon_{\mu \nu \rho \sigma}
\partial_{\mu} A_{\nu}{}^{IJ}\partial_{\rho} A_{\sigma}\ . 
\eea
Unlike the $\omega=\pi/8$ case, when $\omega=0$ the $U(1)$ gauge field
satisfies a Dirichlet boundary condition, and therefore the dual of the
bulk $U(1)$ gauge field is a conserved spin-1 current. According to
the AdS/CFT dictionary, the 3-point correlation function computed
in the $\omega=0$ theory corresponds
to a 3-point function involving two conserved spin-1 currents and a
scalar operator. Similar calculations lead to the 3-point correlation
functions associated with the first and second cubic vertices, which are
given by
\bea
&&\langle {\cO}^{\Delta=2}(\vec x_1)^{IJ}J_i(\vec x_2)^{KL}J_j(\vec x_3)
\rangle\Big|_{\omega=0}=\widetilde{T}^{(1)IJ,KL}_{ij,\Delta=2}\ ,
\nn\\
&&\langle {\cO}^{\Delta=1}(\vec x_1)^{IJ}J_i(\vec x_2)^{KL}J_j(\vec x_3)
\rangle\Big|_{\omega=0}=\widetilde{T}^{(2)IJ,KL}_{ij,\Delta=1}\ ,
\eea
where
\bea
\widetilde{T}^{(1)IJ,KL}_{ij}(\vec x_1,\vec x_2,\vec x_3)&=&
 -\frac{\Gamma[\frac{1+\Delta}2]\Gamma[\frac{\Delta}2]\Gamma[\frac{\Delta}2+1]
 \Gamma[2-\frac{\Delta}2]}{8\sqrt{2}\pi^4\Gamma[\Delta-\frac32]}
(\delta^{IK}\delta^{JL}-\delta^{IL}\delta^{JK})
\nn\\
&&\!\!\!\!\!\!\!\! \!\!\!\!\!
\times\frac{(\Delta-4)J_{ij}(\vec x_2-\vec x_3)+
\Delta J_{ik}(\vec x_2-\vec x_1)J_{kj}(\vec x_1-\vec x_3)}{
|\vec x_2-\vec x_3|^{4-\Delta}|\vec x_1-\vec x_3|^{\Delta}
  |\vec x_1-\vec x_2|^{\Delta}}\ ,
\eea
\bea
&&\widetilde{T}^{(2)IJ,KL}_{ij}(\vec x_1,\vec x_2,\vec x_3)=
-\im\frac{\Gamma[\frac{1+\Delta}2]^2\Gamma
[\frac{\Delta}2+1]\Gamma[\frac{5-\Delta}2]}{2\sqrt{2}\pi^4
\Gamma[\Delta-\frac32]}(\delta^{IK}\delta^{JL}-\delta^{IL}\delta^{JK})
\nn\\
&&\quad \times \frac{|\vec x_1-\vec x_3|^{1-\Delta} \varepsilon_{mkj}}{
 |\vec x_1-\vec x_2|^{\Delta+1}
|\vec x_2-\vec x_3|^{3-\Delta}} J_{ki}(\vec x_2-\vec x_3)
\Big[\frac{(x_3-x_1)_{m}}{
|\vec x_3-\vec x_1|^2}-\frac{(x_3- x_2)_{m}}{|\vec x_3-\vec x_2|^2}
\Big]\ .
\eea
The form of our 3-point correlation function
matches with the general structure of $\langle{\cO}JJ\rangle$ obtained
from a CFT calculation by utilizing
conformal symmetry and current conservation \cite{xiyin}.

Using the lemmata
\bea
\im \varepsilon_{jkm}\frac{\partial}{
\partial {x_{3k}}}{T}^{(1)IJ,KL}_{im,\Delta=1}
(\vec x_1,\vec x_2,\vec x_3)&=&-\widetilde{T}^{(2)IJ,KL}_{ij,\Delta=1}
(\vec x_1,\vec x_2,\vec x_3)\ ,
\nn\\
\im\varepsilon_{jkm}\frac{\partial}{\partial{x_{3k}}}{T}^{(2)IJ,KL}_{im,\Delta=2}(\vec x_1,\vec x_2,\vec x_3)
&=&\widetilde{T}^{(1)IJ,KL}_{ij,\Delta=2}(\vec x_1,\vec x_2,\vec x_3)\ ,
\eea
we find that
\be
\langle {\cO}^{\Delta=1,2}(\vec x_1)^{IJ}J_i(\vec x_2)^{KL}J^{\rm top}_j
(\vec x_3)\rangle\Big|_{\omega=\pi/8}=\langle {\cO}^{\Delta=1,2}
(\vec x_1)^{IJ}J_i(\vec x_2)^{KL}J_j(\vec x_3)\rangle\Big|_{\omega=0}\ ,
\ee
which agrees with (\ref{OOJJ}).

In seeking a holographic CFT interpretation of these results, we know
that the electric-magnetic duality of the $U(1)$ gauge field in the bulk
can be understood in terms of the called $S$-transformation of the
boundary CFT, in which a global $U(1)$ symmetry is gauged and an
off-diagonal Chern-Simons term is added to the CFT \cite{Witten:2003ya}.
Noting that in addition to the $U(N)_k\times U(N)_{-k}$ ABJM model
there also exists an $SU(N)_k\times SU(N)_{-k}$ ABJM model, it is
tempting to interpret the former as the $S$-transform of the latter.
This is motivated by the fact that the bosonic part of the
$U(1)\times U(1)$ sector of the ABJM model has the form
\be
\cL_{\rm ABJM}= -{\rm tr}|(\partial_i-i B_i)C^I|^2+\varepsilon^{ijk}
A_i\partial_j B_k, \quad I=1\cdots 4\ ,
\ee
where $C^I$ are scalar fields in the bi-fundamental representation
of $U(N)\times U(N)$, and $A_i, B_i$ are defined in terms of the
diagonal $U(1)\times U(1)$ gauge fields as
$B_i=A_{1i}-A_{2i}$ and $A_i=A_{1i}+A_{2i}$. However, it has been pointed
out that the bulk dual of the $SU(N)_k\times SU(N)_{-k}$ model is not
simply related to the undeformed $\cN=6$ theory \cite{Aharony:2008gk}.

Turning to the $U(N)_k\times U(N)_{-k}$ ABJM model, in addition to
the $U(1)$ current whose bosonic part is
\be
J_i=\im {\tr}\left(C^{I\star}\overset{\leftrightarrow}{\partial_i}
C^I\right) \ ,
\ee
there also exists a topological current given by
\be
J^{\rm top}_{i}=\varepsilon_{ijk}\partial_j A_k\ .
\ee
The $B_i$ equation implies
\be
J^{\rm top}_i=J_i\ .
\ee
Thus, the relation (\ref{OOJJ}) that we found from the bulk point of view
is manifestly realized due to the on-shell identification
of $J^{\rm top}_i$ with the Noether current $J$ in the
$U(N)_k\times U(N)_{-k}$ ABJM model. Therefore we observe 
that the holographic dual of the $\omega=\pi/8$ theory is not a new
CFT, but instead the ABJM model itself, in the sense that the processes
involving the Noether current $J$ and those involving the
dynamical $U(1)$ in the ABJM model are described by ostensibly distinct
bulk theories with $\omega=0$ and $\omega=\pi/8$ respectively, which,
in turn, are related to each other by electric-magnetic duality.

\section{Conclusions}

   The main goal of this paper has been to initiate an investigation
of correlation functions in the
conformal field theories holographically dual to
the recently discovered $\omega$ deformations of gauged supergravities.
For simplicity, our principal focus has been on the $\cN=6$ supersymmetric
supergravities.  However, since we
obtained these $\omega$-deformed $\cN=6$ theories by truncation from
$\cN=8$, it was of interest to study some of the aspects of the
$\omega$ deformations also in the full $\cN=8$ gauged supergravities.
We therefore also examined the supersymmetry-preserving boundary
conditions in $\omega$-deformed $SO(8)$ gauged $\cN=8$ supergravity.
The inequivalent such theories are characterized by $\omega$ lying in
the interval $[0,\pi/8]$. For any non-vanishing $\omega$ in this
interval we find that consistent Fefferman-Graham boundary conditions
for fluctuations around the trivial AdS$_4$ vacuum can be compatible with
at most an $\cN=3$ subset of the $\cN=8$ supersymmetries.  This result
is obtained under the
assumption that the graviton and gravitini must necessarily obey
Dirichlet boundary conditions, so that they do not correspond to
propagating spin-2 or spin-$3/2$ modes
in the dual boundary theory. In this $\cN=3$ case, all the vectors
must also obey
Dirichlet boundary
conditions. We also find that $\cN=1$ is the maximum allowed supersymmetry
for which some of the vectors can instead obey
Neumann boundary conditions, when $\omega$ is non-vanishing.  Furthermore,
we established that mixed boundary conditions on the vector fields,
where a given vector would have both electric and magnetic components
on the boundary,
are not allowed for any $\cN\ge 1$ supersymmetry.

If $\omega$ does vanish, $\cN=8$ supersymmetry is allowed with
all spin$\geq1$ fields obeying Dirichlet boundary conditions
\cite{Hawking:1983mx}.  We also found that at $\omega=0$
a maximum of $\cN=2$ supersymmetry is compatible with Neumann
boundary conditions imposed on a subset of the vector fields.

The situation as regards supersymmetry-preserving boundary conditions
is very different if we actually truncate the $\cN=8$ theory to
a theory with a lower degree of supersymmetry.
Among such theories, in this paper we have studied the $\omega$-deformed
$SO(6)$ gauged $\cN=6$ supergravity. The undeformed version of this
truncation was studied in \cite{ferrara}.  We constructed
the $\cN=6$ truncation
for the $\omega$-deformed theory, and showed that the $\omega$ deformation
survives, again in the interval $[0,\pi/8]$. We also exhibited the
underlying $SO^*(12)/U(6)$ coset structure of the couplings. Furthermore, we
found that $\cN=6$ supersymmetry-preserving boundary conditions are
possible, provided that $\omega=\pi/8$ (or $\omega=0$). 
When $\omega=\pi/8$, the $SO(6)$ gauge fields still obey
Dirichlet boundary conditions, but the additional $U(1)$ gauge
field obeys a Neumann boundary condition. The $\omega=0$ and
$\omega=\pi/8$ theories are related by a $U(1)$ electric-magnetic duality.

   The $U(1)$ gauge field appears in the equations of motion only through
its field strength, since none of the $\cN=6$ fields are charged
under the $U(1)$.  We showed that the embedding of the $\omega$-deformed 
$\cN=6$ theories into IIA supergravity reduced on $\CP^3$ can be
straightforwardly accomplished, in view of the fact that
Kaluza-Klein reductions on spheres or other curved manifolds are
necessarily performed at the level of the equations of motion,
as explained in more detail in appendix A.

  We computed the leading-order examples of
$\omega$-dependent tree-level amplitudes in the $\cN=6$ theories, as a step
towards understanding the $\omega$ deformation from a dual holographic
viewpoint. Up to 3-point tree-level graphs, we found that the only
$\omega$-dependent amplitudes are those involving the trilinear
coupling of a
$SO(6)$ gauge field with the $U(1)$ gauge field and a scalar or
pseudoscalar field. We computed these amplitudes, which are
parity-violating. The amplitudes turned out to be finite without the need
for any regularisation.  These results would also have a wider applicability
in other situations where one has gauge fields obeying Neumann boundary
conditions as well as gauge fields obeying Dirichlet boundary conditions.

    We also computed the associated correlation functions in the
undeformed theory. Inspired by Witten's holographic
 interpretation of bulk electric-magnetic duality \cite{Witten:2003ya}, we
found that the electric-magnetic duality transformation of
the $U(1)$ field that is required when relating the $\omega=0$ and
$\omega=\pi/8$ bulk theories has the effect of
interchanging the Noether current and topological current in the
amplitudes of the 
$U(1)\times U(1)$ sector of the $U(N)_k\times U(N)_{-k}$ ABJM model.

   Although we focused on computing the amplitudes in 
the $\omega$-deformed $SO(6)$ gauged supergravity, it would
   be interesting to study other $\omega$-deformed gauged supergravities 
that admit an AdS$_4$ vacuum, in framework of the AdS/CFT
   correspondence. At present, the following gauged $\cN=8$ 
supergravity theories are known to admit $\omega$ deformations with 
   supersymmetric AdS vacua \cite{Gallerati:2014xra}:                                                                                                 
   \begin{itemize}                                                                                                                               
     \item $SO(8)$ supergravity with $\cN=8$ supersymmetry.                                                                                      
     \item $SO(1,7)$ and $\left[SO(1,1)\times SO(6)\right]\ltimes T^{12}$ supergravities with $\cN=4$ supersymmetry.                             
     \item $SO(8)$, $SO(1,7)$ and $ISO(1,7)$ gauged supergravities with $\cN=3$ supersymmetry.                                                   
   \end{itemize}                                                                                                                                 
   We have shown in this paper that in the $SO(8)$ gauged 
$\cN=8$ theory, the boundary conditions
   preserve at most $\cN=3$ supersymmetry for non-vanishing $\omega$, 
even though the vacuum itself preserves $\cN=8$ supersymmetry.
   The supersymmetry-preserving boundary conditions and holographic aspects for the remaining 
   supersymmetric vacua deserve further investigation.

\vskip 1in
{\noindent\large  \bf Acknowledgments}
\vskip 0.1in

  We are grateful to Daniel Jafferis, Hermann Nicolai,
  Gianluca Inverso, Xi Yin for helpful discussions.
  Y.P. is grateful to the Mainz Institute for Theoretical Physics (MITP)
for its hospitality and its partial support during the completion of this work.
A.B. is grateful to the Mitchell Institute for hospitality during an
extended visit to Texas A\&M University.
The work of C.N.P. is supported in part by DOE grant DE-FG02-13ER42020. The
work of E.S. is supported in part by NSF grant PHY-1214344.

\newpage
\begin{appendix}

\section{$\omega$-Deformed $\cN=6$ Supergravity and Higher Dimensions}


An outstanding problem is to find whether the $\omega$-deformed supergravities
have any higher-dimensional origin. In the case of $\omega$-deformed
$\cN=8$ supergravity, it has been suggested that to embed it into
eleven-dimensional supergravity would probably require first extending
the eleven-dimensional theory to some kind of a doubled theory
include a ``dual graviton'' in addition to the usual one \cite{dewitnicnew}.
The idea, essentially, is that the $\omega$ deformation amounts to a gauging
in which some combination of the 28 dual vector gauge fields as well as the
28 original gauge fields of the four-dimensional theory would participate
in the minimal couplings to the other fields of the $\cN=8$ multiplet, and
such gaugings could not arise unless the dual fields were already
themselves embedded into the eleven-dimensional theory, as components of
a dual graviton.
However, the construction of such a doubled eleven-dimensional
theory, with non-linear couplings for the dual gravitons,
remains an open problem.

The situation is rather different in the case of the
$\omega$-deformed $\cN=6$ supergravities.  As we discussed in section 3,
at the level of the four-dimensional equations of motion the
$\omega$ parameter in the deformed $\cN=6$ theories can be absorbed by
means of a duality transformation of the $U(1)$ gauge field.  This
can be done because unlike the $SO(6)$ gauge fields, which
have minimal couplings to other fields in supermultiplet,
the $U(1)$ gauge field has no minimal couplings, and it enters the
equations of motion purely through its field strength.  The non-triviality
of the $\omega$ parameter in the $\cN=6$ theory stems solely from the
fact that it cannot be absorbed by any local field redefinition at the
level of the Lagrangian, and thus it can affect quantum properties of
the theory (such as correlation functions in the dual theory).

   For the above reasons, the question of whether the $\omega$-deformed
$\cN=6$ supergravities can be embedded in a higher-dimensional theory is
rather different from the $\cN=8$ case.  First of all, we note that the highly
non-trivial mechanism whereby a sphere reduction can give rise to a
consistent truncation in the lower dimension is one that always operates
at the level of the equations of motion, rather than at the level of the
Lagrangian.  In other words, the lower-dimensional theory, such as
the standard $\cN=8$ gauged supergravity, emerges at the level of the
equations of motion when the reduction ansatz is substituted into the
higher-dimensional equations of motion.  One cannot instead substitute
into the higher-dimensional action and thereby obtain the lower-dimensional
action.  An illustration, pertinent to our present discussion,
of why this is the case is provided by a very
early example of a non-trivial consistent sphere reduction that was
obtained in \cite{popecon}.  In that paper it was shown that the
four-dimensional Einstein-Maxwell theory with a negative cosmological
constant could be consistently embedded in eleven-dimensional
supergravity, whose bosonic Lagrangian is ${\cal L}_{11}= \hat R\,
\hat*\oneone -\ft12 {\hat *\hat F}_\4\wedge \hat F_\4 +
 \ft16 \hat F_\4\wedge \hat F_\4\wedge \hat A_\3$, with the reduction
ansatz being given by \cite{popecon}
\bea
d\hat s_{11}^2 &=& ds_4^2 + (d\psi + A+B)^2 + d\Sigma_6^2\ ,\nn\\
\hat F_\4 &=& 6 m \epsilon_\4  -{*F}\wedge J\ .\label{d11ans}
\eea
Here $ds_4^2$ is the four-dimensional metric, with volume form $\epsilon_\4$,
$d\Sigma_6^2$ is the Fubini-Study metric on $\CP^3$ with Ricci tensor
$R_{ab} = 8 m^2 g_{ab}$, $dB=2 m J$ where $J$ is the K\"ahler form of
$\CP^3$, and $*$ denotes the four-dimensional Hodge dual.  This reduction
ansatz obeys the eleven-dimensional equations of motion, and the Bianchi
identity $d\hat F_\4=0$, if and only if the four-dimensional fields
$g_{\mu\nu}$ and $A_\mu$ satisfy the Einstein-Maxwell equations \cite{popecon}
\be
R_{\mu\nu} = 2 (F_{\mu\rho} F_{\nu}{}^\rho - \ft14 F^2\, g_{\mu\nu})
   -12 m^2 g_{\mu\nu}\ ,\qquad \nabla_\mu F^{\mu\nu}=0\ .
\ee
The fact that the ansatz for the eleven-dimensional 4-form in (\ref{d11ans})
obeys the
Bianchi identity $d\hat F_\4=0$ only upon the use of the four-dimensional
equations of motion illustrates the fact that one could not write down
a reduction ansatz on the original fundamental fields $\hat g_{MN}$ and
$\hat A_\3$ appearing in the
eleven-dimensional Lagrangian.

   The embedding of the Einstein-Maxwell theory given by (\ref{d11ans}) is
in fact itself a consistent truncation of the embedding of the (bosonic
sector) of the standard
gauged $\cN=6$ supergravity into eleven dimensions, namely where the
fields are truncated to the $SO(6)$ singlets.  The Maxwell field in
(\ref{d11ans}) is precisely the $U(1)$ gauge field of the $\cN=6$ theory.

  The question of whether one can embed the $\omega$-deformed family
of $\cN=6$ gauged supergravities in eleven-dimensional supergravity is
a slightly tricky one.  Since the consistent reduction must be performed
at the level of the equations of motion, and since the $\omega$ parameter
in the gauged $\cN=6$ supergravities can be absorbed by means of
local scalar field redefinitions and a $U(1)$ duality transformation at
the level of the equations of motion, it follows that the entire
family of $\omega$-deformed theories can be embedded into eleven-dimensional
supergravity.  Of course, since the $U(1)$ gauge potentials for two
different values of $\omega$ are non-locally related, and since the bare
$U(1)$ gauge potential appears in the metric ansatz, as in the
further truncation in (\ref{d11ans}),
this means that the eleven-dimensional embeddings for two different
values of $\omega$ would be non-locally related.

  One can instead consider the embedding  the
$\omega$-deformed $\cN=6$ supergravities into
the ten-dimensional type IIA supergravity.  As was shown in
\cite{nilpop}, if one makes an $S^1$ reduction on the Hopf fibres of
the $S^7$ embedding of $\cN=8$ supergravity into $D=11$, truncating
to the $U(1)$ singlets, one obtains the $\cN=6$ gauged supergravity as
a consistent reduction of type IIA supergravity on $\CP^3$.  In particular,
the $U(1)$ gauge field in the $\cN=6$ theory is now coming not from the
ten-dimensional metric, but rather, it is the Ramond-Ramond 2-form field
strength already present in the type IIA theory.  (Which arose, of course,
from the Kaluza-Klein vector of the $S^1$ reduction from $D=11$.)  Thus
in the embedding of the $\cN=6$ theory into the type IIA theory the
reduction ansatz only requires the knowledge of the $U(1)$
{\it field strength}, and not its 1-form potential.  Accordingly,
not only can one embed any of the $\omega$-deformed $\cN=6$ supergravities
into ten-dimensional type IIA supergravity, but also the relation between
the embeddings for two different values of $\omega$ can now be
expressed purely locally, since the $U(1)$ gauge potential does not
appear in the reduction ansatz from $D=10$ to $D=4$.

\section{Supersymmetric Boundary Conditions in $\omega$-deformed ${\cal N}=8$ Theory}

 It was shown long ago that in the de Wit-Nicolai theory, there exist
${\cal N}=8$ supersymmetry-preserving boundary conditions
\cite{Hawking:1983mx}. Explicitly, these boundary conditions, with
the gauge choices (\ref{gaugechoices}), are given by\footnote{In this
appendix we shall use $I,J,\ldots$ to denote $SO(8)$ indices, while
$i,j,\ldots$ denote boundary coordinate indices.}
\be
e^{\hat{r}}_{(0)i}=\delta^{\hat{r}}_{i},\quad\psi^{I}_{(0)i}=0,
\quad A^{IJ}_{(0)i}=0,\quad\chi^{IJK}=0,\quad {\cal S}^{IJKL}_{(2)}=0,
\quad  {\cal P}^{IJKL}_{(1)}=0\ .
\ee
In this appendix, we shall show that in the $\omega$-deformed ${\cal N}=8$
theories one can no longer impose consistent boundary
conditions that preserve the full ${\cal N}=8$ supersymmetry.
Note that we are making the key assumption in this analysis that the
fields of spins 2 and $3/2$ must
obey Dirichlet boundary conditions, to avoid having propagating graviton
or gravitino modes in the boundary theory.  The goal in this appendix is
then to determine the possible numbers of
supersymmetries
that {\it are} preserved by appropriate choices of boundary conditions in
the full $\omega$-deformed ${\cal N}=8$ theory.

Because shall be considering a subset ${\cal N}<8$ of the full ${\cal N}=8$
supersymmetry of the theory itself, the supersymmetry variations of
the boundary conditions on the gravitini
will only determine the boundary conditions on a subset of the vector fields.
For the remaining vector fields, we shall begin by considering the cases of
purely
Dirichlet or Neumann boundary conditions on these fields. Later on, when
considering
the $\cN=1$ case, we shall allow for all possible mixed boundary conditions
on the vector fields. We shall show, however,  that such
mixed boundary conditions are ruled out.  Applying similar considerations,
we then show that
mixed boundary conditions on vectors are not possible for any $\cN>1$ either,
thus justifying our previous restriction to the purely Dirichlet or Neumann
possibilities.

\subsection{${\cal N}\geq4$ Supersymmetry.}

We shall first show that in the $\omega$-deformed ${\cal N}=8$ theory there
are no boundary conditions that preserve an $\cN=4$ subset of the
${\cal N}=8$ supersymmetries. In turn, this then implies that there
cannot be boundary conditions preserving any number ${\cal N}\ge4$ of the
supersymmetries, either.

To show that ${\cal N}=4$ supersymmetry-preserving boundary conditions are
not possible when $\omega\ne0$, we split the $SO(8)$ indices into
$I,J=1,2,3,4$, and $r,s=5,...8$.  It suffices in this ${\cal N}=4$ case
to consider the boundary
conditions on the vector fields $A_i^{rs}$.  Consider first the case
when they obey Dirichlet boundary conditions, i.e. $A_{(0)i}^{rs}=0$.
The supersymmetry
variation of this boundary condition, under the ${\cal N}=4$
subset $\epsilon^I$ of the supersymmetry parameters
(i.e. with $\epsilon^r=0$), leads to
\be
\cos2\omega\  {\cal P}^{rsIJ}_{(1)} +\sin2\omega\ {\cal S}_{(1)}^{rsIJ}=0,\quad 
\sin2\omega\  {\cal P}^{rsIJ}_{(2)}-\cos2\omega\ {\cal S}_{(2)}^{rsIJ}=0\ .
\ee
It then follows from the $SU8)$ duality condition (\ref{su8duality2}) that
\be
\cos2\omega\  {\cal P}^{rsIJ}_{(1)} -\sin2\omega\ {\cal S}_{(1)}^{rsIJ}=0,\quad 
\sin2\omega\  {\cal P}^{rsIJ}_{(2)}+\cos2\omega\ {\cal S}_{(2)}^{rsIJ}=0\ .
\ee
Therefore, if $\omega\neq 0$, then we find ${\cal S}_{(1)}^{rsIJ}=
{\cal P}_{(1)}^{rsIJ}={\cal S}_{(2)}^{rsIJ}={\cal P}_{(2)}^{rsIJ}=0$,
which are not acceptable as boundary conditions since they would imply
that these scalars all vanished everywhere.

 Now consider instead imposing Neumann boundary conditions on $A^{rs}_i$.
If these boundary conditions preserved ${\cal N}=4$ supersymmetry,
they would
 also imply the existence of such boundary conditions that preserved
${\cal N}=3$ supersymmetry.  However, as we shall show explicitly in the
next subsection,  there cannot exist ${\cal N}=3$ boundary conditions
with $A^{rs}_i$ satisfying Neumann boundary condition.  Thus we have
established that when $\omega\ne0$ there can exist no choice of boundary
conditions that preserves ${\cal N}\ge 4$ supersymmetry.

\subsection{$\cN = 3$ Supersymmetry.}

In the ${\cal N}=3$ case, we decompose the $SO(8)$ indices so that
$I,J=1,2,3$, and $r,s=4,5,...8$.  As we shall see below, there do in fact
exist consistent ${\cal N}=3$ boundary conditions for $\omega\ne0$, and, in
particular, we can no longer derive an immediate contradiction, as we did
for ${\cal N}=4$, merely by considering the restrictions following from
imposing Dirichlet or Neumann boundary conditions on $A_i^{rs}$. Instead,
we begin the analysis here by noting that
the supersymmetry variation of the Dirichlet boundary condition on the
gravitini requires
\be
A^{rI}_{(0)i}=0,\qquad A^{IJ}_{(0)i}=0\ .
\ee
The vanishing of the supersymmetry variations of $A^{rI}_{(0)i}$ and
$A^{IJ}_{(0)i}$ then imply
\be
\cos2\omega\  {\chi}^{rIJ}_++\im \sin2\omega \gamma_5{\chi}^{rIJ}_-=0,\quad \cos2\omega\  {\chi}^{IJK}_+
+\im \sin2\omega\ \gamma_5{\chi}^{IJK}_-=0\ .
\label{N=3bnd1}
\ee
The second equation is automatically invariant under ${\cal N}=3$
supersymmetry variations. Demanding that the first equation be invariant
under
\bea
\delta\chi^{rIJ}_{+} &=& -{\cal S}^{rIJK}_{(2)}{\epsilon}^{K}_+
+2\im{\cal P}^{rstI}_{(1)}\gamma_5{\epsilon}^{I}_--{\rm i}\slashed{D}{\cal P}_{(1)}^{rIJK}\gamma_5{\epsilon}^{K}_+\ ,
\nn\w2
\delta{\chi}^{rIJ}_{-} &=& 2{\cal S}^{rIJK}_{(1)}{\epsilon}^{K}_--{\rm i}{\cal P}^{rIJK}_{(2)}\gamma_5{\epsilon}^{K}_+
+\slashed{D}{\cal S}_{(1)}^{rIJK}{\epsilon}^{K}_+\ ,
\eea
leads to
\be
\cos2\omega\  {\cal P}^{rIJK}_{(1)}+\sin2\omega\ {\cal S}_{(1)}^{rIJK}=0,\quad 
\sin2\omega\  {\cal P}^{rIJK}_{(2)}-\cos2\omega\ {\cal S}_{(2)}^{rIJK}=0\ .
\label{N=3bnd2}
\ee
Using the duality property (\ref{su8duality2}) of the scalars, the above
equations imply that
\be
\cos2\omega\  {\cal P}^{rstp}_{(1)}-\sin2\omega\ {\cal S}_{(1)}^{rstp}=0,\quad \sin2\omega\  {\cal P}^{rstp}_{(2)}
+\cos2\omega\ {\cal S}_{(2)}^{rstp}=0\ .
\label{N=3bnd22}
\ee
It can be verified that the first equation in (\ref{N=3bnd1}) guarantees
that equations (\ref{N=3bnd2}) and (\ref{N=3bnd22}) are invariant under
the ${\cal N}=3$ supersymmetry variations.

  We now consider the possible boundary conditions on the vector
fields $A_i^{rs}$.  Let us first consider imposing Dirichlet boundary
conditions on $A_i^{rs}$. The vanishing of the variation of
$A^{rs}_{(0)i}$ requires
\be
\cos2\omega\  {\chi}^{rsI}_++\im \sin2\omega\ \gamma_5{\chi}^{rsI}_- =0\ .
\label{N=3bnd3}
\ee
The supersymmetry variation of this condition, using
\bea
\delta\chi^{rsI}_{+} &=& -{\cal S}^{rsIJ}_{(2)}{\epsilon}^{J}_++2\im{\cal P}^{rsIJ}_{(1)}\gamma_5{\epsilon}^{J}_-
-{\rm i}\slashed{D}{\cal P}_{(1)}^{rsIJ}\gamma_5{\epsilon}^{J}_+\ ,
\nn\w2
\delta{\chi}^{rsJ}_{-} &=& 2{\cal S}^{rsIJ}_{(1)}{\epsilon}^{J}_--{\rm i}{\cal P}^{rsIJ}_{(2)}\gamma_5{\epsilon}^{J}_+
+\slashed{D}{\cal S}_{(1)}^{rsIJ}{\epsilon}^{J}_+\ ,
\eea
implies
\be
\cos2\omega\  {\cal P}^{rsIJ}_{(1)}+\sin2\omega\ {\cal S}_{(1)}^{rsIJ}=0,\quad 
\sin2\omega\  {\cal P}^{rsIJ}_{(2)}-\cos2\omega\ {\cal S}_{(2)}^{rsIJ}=0\ .
\label{N=3bnd4}
\ee
 Using the duality property (\ref{su8duality2}) of the scalars,
(\ref{N=3bnd4}) implies
\be
\cos2\omega\  {\cal P}^{Irst}_{(1)}-\sin2\omega\ {\cal S}_{(1)}^{Irst}=0,\quad 
\sin2\omega\  {\cal P}^{Irst}_{(2)}+\cos2\omega\ {\cal S}_{(2)}^{Irst}=0\ .
\label{N=3bnd5}
\ee
To preserve these conditions under the ${\cal N}=3$ supersymmetry variations
\bea
\delta{\cal S}^{Irst}_{(1)} &=& \bar{{\epsilon}}^{I}_+{\chi}_{-}^{rst} -\ft1{2}\varepsilon^{IJK}\epsilon^{rstpq}\bar{\epsilon}^{J}_+{\chi}_{-}^{Kpq}\ ,
\nn\w2
\delta{\cal P}^{Irst}_{(1)} &=& -\im\Big(\bar{\epsilon}^{I}_+\gamma_5{\chi}_{+}^{rst}
+\ft1{2}\varepsilon^{IJK}\epsilon^{rstpq}\bar{\epsilon}^{J}_+\gamma_5\widetilde{\chi}_{+}^{Kpq}\Big)\ ,
\nn\w2
\delta{\cal S}^{Irst}_{(2)} &=& \Big(\bar{\epsilon}^{I}_-{\chi}_{+}^{rst}
-\ft1{2}\varepsilon^{IJK}\epsilon^{rstpq}\bar{\epsilon}^{J}_-{\chi}_{+}^{Kpq}+\bar{\epsilon}^{I}_+\slashed{D}{\chi}_{+}^{rst}
-\ft1{2}\varepsilon^{IJK}\epsilon^{rstpq}\bar{\epsilon}^{J}_+\slashed{D}{\chi}_{+}^{Kpq}\Big)\ ,
\w2
\delta{\cal P}^{Irst}_{(2)} &=& -\im\Big(\bar{\epsilon}^{I}_-\gamma_5{\chi}_{-}^{rst}
+\ft1{2}\varepsilon^{IJK}\epsilon^{rstpq}\bar{\epsilon}^{J}_-\gamma_5{\chi}_{-}^{Kpq}-\bar{\epsilon}^{I}_+\gamma_5\slashed{D}{\chi}_{-}^{rst]}
-\ft1{2}\varepsilon^{IJK}\epsilon^{rstpq}\bar{\epsilon}^{J}_+\gamma_5\slashed{D}{\chi}_{-}^{Kpq}\Big)\ ,
\nn
\eea
we need to impose
\be
\cos2\omega\  {\chi}^{rst}_+-\im \sin2\omega\ \gamma_5{\chi}^{rst}_-=0\ .
\label{N=3bnd6}
\ee
The condition (\ref{N=3bnd6}) is preserved under the supersymmetry variations
\bea
\delta\chi^{rst}_{+}&=&-{\cal S}^{rstI}_{(2)}{\epsilon}^{I}_++2\im{\cal P}^{rstI}_{(1)}\gamma_5{\epsilon}^{I}_--{\rm i}\slashed{D}{\cal P}_{(1)}^{rstI}\gamma_5{\epsilon}^{I}_+\ ,
\nn\\
\delta{\chi}^{rst}_{-}&=&2{\cal S}^{rstI}_{(1)}{\epsilon}^{I}_--{\rm i}{\cal P}^{rstI}_{(2)}\gamma_5{\epsilon}^{I}_++\slashed{D}{\cal S}_{(1)}^{rstI}{\epsilon}^{I}_+\ ,
\nn\\
\eea
as a consequence of (\ref{N=3bnd5}).

In summary, we have found a consistent set of ${\cal N}=3$
supersymmetry-preserving boundary conditions, which takes the form
\bea
&&A^{rs}_{(0)i}=0,\qquad A^{rI}_{(0)i}=0,\qquad A^{IJ}_{(0)}=0\ ,
\nn\\
&&\cos2\omega\  {\chi}^{rIJ}_++\im \sin2\omega\ \gamma_5{\chi}^{rIJ}_-=0,\quad \cos2\omega\  {\chi}^{IJK}_+
+\im \sin2\omega\ \gamma_5{\chi}^{IJK}_-=0\ ,
\nn\\
&&\cos2\omega\  {\chi}^{rsI}_++\im \sin2\omega\ \gamma_5{\chi}^{rsI}_-=0,\quad \cos2\omega\  {\chi}^{rst}_+
-\im \sin2\omega\ \gamma_5{\chi}^{rst}_-=0\ ,
\nn\\
&&\cos2\omega\  {\cal P}^{Irst}_{(1)}-\sin2\omega\ {\cal S}_{(1)}^{Irst}=0\ ,\quad 
\sin2\omega\  {\cal P}^{Irst}_{(2)}+\cos2\omega\ {\cal S}_{(2)}^{Irst}=0\ ,
\nn\\
&&\cos2\omega\  {\cal P}^{rIJK}_{(1)}+\sin2\omega\ {\cal S}_{(1)}^{rIJK}=0\ ,\quad 
\sin2\omega\  {\cal P}^{rIJK}_{(2)}-\cos2\omega\ {\cal S}_{(2)}^{rIJK}=0\ .
\label{totN=3bnd}
\eea
Equations (\ref{N=3bnd22}) and (\ref{N=3bnd4}) are implied by the duals of
the last two equations in (\ref{totN=3bnd}).

We shall now show that it is not possible to impose instead Neumann boundary
conditions on $A^{rs}_i$ while preserving ${\cal N}=3$ supersymmetry.
Imposing  the Neumann boundary condition $A^{rs}_{(1)i}=0$, its
supersymmetry variation under
\bea
\delta A^{rs}_{(1)i}&=&-\Big({\cal S}_{(1)}^{rspq}\bar{\epsilon}^{I}_+\gamma_{(0)i}{\chi}^{pqI}_{+}
+\im{\cal P}_{(1)}^{rspq}\bar{\epsilon}^{I}_+\gamma_{(0)i}\gamma_5{\chi}^{pqI}_{-}
\nn\w2
&&\qquad +{\cal S}_{(1)}^{rsIJ}\bar{\epsilon}^{K}_+\gamma_{(0)i}{\chi}^{IJK}_{+}
+\im{\cal P}_{(1)}^{rsIJ}\bar{\epsilon}^{K}_+\gamma_{(0)i}\gamma_5{\chi}^{IJK}_{-}
\nn\\
&&\qquad+2{\cal S}_{(1)}^{rstI}\bar{\epsilon}^{J}_+\gamma_{(0)i}{\chi}^{tIJ}_{+}
+\im 2{\cal P}_{(1)}^{rstI}\bar{\epsilon}^{J}_+\gamma_{(0)i}\gamma_5{\chi}^{tIJ}_{-}
\nn\\
&&\qquad -D_{i}(A_{(0)})(\cos2\omega\ \bar{\epsilon}^{I}_{+}{\chi}^{rsI}_{-}
+{\rm i}\sin2\omega\ \bar{\epsilon}^{I}_{+}\gamma_5{\chi}^{rsI}_+)\Big)\ ,
\eea
requires
\be
\sin2\omega\  {\chi}^{rsI}_+-\im \cos2\omega\ \gamma_5{\chi}^{rsI}_-=0\ .
\label{N=3bnd7}
\ee
Furthermore, utilizing (\ref{N=3bnd22}), one can see that the vanishing of $A^{rs}_{(1)i}$ also requires
\be
\cos2\omega\ \chi_+^{rsI}+\im\sin2\omega\ \gamma_5\chi^{rsI}_-=0\ .
\ee
The above equation, together with the second equation in (\ref{N=3bnd7}),
leads to
\be
\chi_+^{rsI}=0,\quad \chi_-^{rsI}=0\ .
\ee
This is too strong a condition on the fermions $\chi^{rsI}$, since it
implies that they vanish everywhere.  Thus, we find that there cannot exist
any consistent ${\cal N}=3$ boundary condition in which $A^{rs}_i$
satisfies Neumann boundary conditions.

\subsection{$\cN=2$ Supersymmetry}

   There can clearly exist ${\cal N}=2$ boundary conditions that simply
follow as a reduction of the ${\cal N}=3$ boundary conditions that we
obtained above.  However, since ${\cal N}=2$ supersymmetry is less
restrictive than ${\cal N}=3$, there could also exist further possible
boundary conditions that are compatible with ${\cal N}=2$ but not with
${\cal N}=3$.  We shall therefore now
proceed to the analyze the case with ${\cal N}=2$ supersymmetry. We
introduce indices $I,J=1,2$, and let $r,s$ range from 3 to 7.
The Dirichlet boundary conditions for the gravitini imply
\bea
{\psi}^I_{(0)i}&=&0 \quad \Rightarrow \quad A^{IJ}_{(0)i}=0,\nn\\
{\psi}^r_{(0)i}&=&0 \quad \Rightarrow \quad A^{rI}_{(0)i}=0 \qquad  \Rightarrow \quad \cos2\omega\  {\chi}^{rIJ}_+
+\im \sin2\omega\ \gamma_5{\chi}^{rIJ}_-=0\ .
\eea
We now follow a sequence of steps paralleling those that we used for
the ${\cal N}=3$ case. Using the supersymmetry variations of the
leading terms in the Fefferman-Graham expansions for the spin-1,
spin-$\ft12$ and spin-0 fields, namely
\allowdisplaybreaks
\bea
\delta{\chi}^{rIJ}_{+} &=& 0, \quad \delta{\chi}^{rIJ}_{-}=0\ ,
\nn\w2
\delta{\chi}^{rsI}_{+} &=& -{\cal S}^{rsIJ}_{(2)}{\epsilon}^{J}_++2{\rm i}{\cal P}^{rsIJ}_{(1)}\gamma_5{\epsilon}^{J}_-
- {\rm i}\slashed{D}{\cal P}_{(1)}^{rsIJ}\gamma_5{\epsilon}^{J}_+\ ,
\nn\w2
\delta{\chi}^{rsI}_{-} &=& 2{\cal S}^{rsIJ}_{(1)}{\epsilon}^{J}_--{\rm i}{\cal P}^{rsIJ}_{(2)}\gamma_5{\epsilon}^{J}_+
+\slashed{D}{\cal S}_{(1)}^{rsIJ}{\epsilon}^{J}_+\ ,
\nn\w2
\delta{\chi}^{rst}_{+} &=& -{\cal S}^{rstI}_{(2)}{\epsilon}^{I}_+ +2{\rm i}{\cal P}^{rstI}_{(1)}\gamma_5{\epsilon}^{I}_-
-{\rm i}\slashed{D}{\cal P}_{(1)}^{rstI}\gamma_5{\epsilon}^{I}_+\ ,
\nn\w2
\delta{\chi}^{rst}_{-} &=& 2{\cal S}^{rstI}_{(1)}{\epsilon}^{I}_--{\rm i}{\cal P}^{rstI}_{(2)}\gamma_5{\epsilon}^{I}_+
+\slashed{D}{\cal S}_{(1)}^{rstI}{\epsilon}^{I}_+\ ,
\nn\w2
\delta{\cal S}^{Irst}_{(1)} &=& \Big(\bar{{\epsilon}}^{I}_+{\chi}_{-}^{rst}
- \ft1{3!}\varepsilon^{rstpqu} \epsilon^{IJ} \bar{{\epsilon}}^{J}_+{\chi}_{-}^{pqu}\Big)\ ,
\nn\w2
\delta{\cal P}^{Irst}_{(1)} &=& -\im\Big(\bar{{\epsilon}}^{I}_+\gamma_5{\chi}_{+}^{rst}
+ \ft1{3!}\varepsilon^{rstpqu} \epsilon^{IJ}\bar{{\epsilon}}^{J}_+\gamma_5{\chi}_{+}^{pqu}\Big)\ ,
\nn\w2
\delta{\cal S}^{Irst}_{(2)} &=& \Big(\bar{{\epsilon}}^{I}_-{\chi}_{+}^{rst}
- \ft1{3!}\varepsilon^{rstpqu} \epsilon^{IJ}\bar{{\epsilon}}^{J}_-{\chi}_{+}^{pqu}
\nn\w2
&&+\bar{{\epsilon}}^{I}_+\slashed{D}{\chi}_{+}^{rst}
- \ft1{3!}\varepsilon^{rstpqu} \epsilon^{IJ} \bar{{\epsilon}}^{J}_+\slashed{D}{\chi}_{+}^{pqu}\Big)\ ,
\nn\w2
\delta{\cal P}^{Irst}_{(2)} &=& -\im\Big(\bar{{\epsilon}}^{I}_-\gamma_5{\chi}_{-}^{rst}
+\ft1{3!}\varepsilon^{rstpqu} \epsilon^{IJ}\bar{{\epsilon}}^{J}_-\gamma_5{\chi}_{-}^{pqu}
\nn\w2
&&-\bar{{\epsilon}}^{I}_+\gamma_5\slashed{D}{\chi}_{-}^{rst]}
- \ft1{3!}\varepsilon^{rstpqu} \epsilon^{IJ}\bar{{\epsilon}}^{J}_+\gamma_5\slashed{D}{\chi}_{-}^{pqu}\Big)\ ,
\nn\w2
\delta{\cal S}^{IJrs}_{(1)} &=& 2\bar{{\epsilon}}^{[I}_+{\chi}_{-}^{J]rs},\quad
\delta{\cal P}^{IJrs}_{(1)} = -2\im\bar{{\epsilon}}^{[I}_+\gamma_5{\chi}_{+}^{J]rs}\ ,
\nn\w2
\delta{\cal S}^{IJrs}_{(2)}&=&2\Big(\bar{{\epsilon}}^{[I}_-{\chi}_{+}^{J]rs} +\bar{{\epsilon}}^{[I}_+\slashed{D}{\chi}_{+}^{J]rs}\Big)\ ,
\nn\w2
\delta{\cal P}^{IJrs}_{(2)} &=& -2\im\Big(\bar{{\epsilon}}^{[I}_-\gamma_5{\chi}_{-}^{J]rs}
-\bar{{\epsilon}}^{[I}_+\gamma_5\slashed{D}{\chi}_{-}^{J]rs}\Big)\ ,
\nn\w2
\delta A^{rs}_{(0)i} &=& -\Big(\cos2\omega\ {\epsilon}^{I}_+\gamma_{(0)i}{\chi}^{rsI}_{+}+{\rm i}\sin2\omega\ {\epsilon}^{I}_+\gamma_{(0)i}\gamma_5{\chi}^{rsI}_{-}\Big)\ ,
\nn\w2
\delta A^{rs}_{(1)i} &=& -\Big({\cal S}_{(1)}^{rstp}{\epsilon}^{I}_+\gamma_{(0)i}{\chi}^{tpI}_{+}
+\im{\cal P}_{(1)}^{rstp}{\epsilon}^{I}_+\gamma_{(0)i}\gamma_5{\chi}^{tpI}_{-}
\nn\w2
&& +2{\cal S}_{(1)}^{rstI}{\epsilon}^{J}_+\gamma_{(0)i}{\chi}^{tIJ}_{+}
+2\im{\cal P}_{(1)}^{rstI}{\epsilon}^{J}_+\gamma_{(0)i}\gamma_5{\chi}^{tIJ}_{-}
\nn\w2
&& -D_{i}(A_{(0)})(\cos2\omega\ {\epsilon}^{I}_{+}{\chi}^{rsI}_{-}+{\rm i}\sin2\omega\ {\epsilon}^{I}_{+}\gamma_5{\chi}^{rsI}_+)\Big)\ ,
\eea
we find that we can obtain consistent ${\cal N}=2$ boundary conditions in
which all the spin-1 fields satisfy Dirichlet boundary conditions.  The
full set of boundary conditions in this case is given by
\bea
&&A^{rI}_{(0)i}=0,\quad A^{IJ}_{(0)i}=0,\quad A^{rs}_{(0)i}=0,\quad \alpha_{rst}{\chi}^{rst}_++\im\gamma_5\beta_{rst}{\chi}^{rst}_-=0,\nn\w2
&&\cos2\omega\ \  {\chi}^{rIJ}_++\im \sin2\omega\ \ \gamma_5{\chi}^{rIJ}_-
=0,\quad \cos 2\omega\ {\chi}^{rsI}_++\im \sin 2\omega\ \gamma_5{\chi}^{rsI}_-=0\ ,
\nn\w2
&&\sin2\omega\ \  {\cal S}^{rsIJ}_{(1)}+\cos2\omega\ \ {\cal P}^{rsIJ}_{(1)}=0\ , \quad
\cos2\omega\ \  {\cal S}^{rsIJ}_{(2)}-\sin2\omega\ \ {\cal P}^{rsIJ}_{(2)}=0\ ,\nn\w2
&&\beta_{rst} {\cal S}^{rstI}_{(1)}+\alpha_{rst} {\cal P}^{rstI}_{(1)}=0\ , \quad
\alpha_{rst}   {\cal S}^{rstI}_{(2)}-\beta_{rst} {\cal P}^{rstI}_{(2)}=0,
\label{xxxx}
\eea
where the indices $r,s,t$ are not summed, and the coefficients
$\alpha_{rst}$ and $\beta_{rst}$ are constants.  The point here is that
the leading-order terms in the Fefferman-Graham expansions of the
fermions $\chi^{rst}$,
together with those of the scalars $\cS^{rstI}$ and $\cP^{rstI}$, form
a closed multiplet, whose supersymmetry variations are not related to the
expansions of any other fields, and so we are free to impose whatever
self-consistent boundary conditions we wish on these fields.  We can make
an independent choice of boundary condition for each independent
component of $\chi^{rst}$, with the boundary conditions for the
corresponding $\cS^{rstI}$ and $\cP^{rstI}$ fields then following from
the ${\cal N}=2$ supersymmetry.  In view of the antisymmetry and
duality properties of $\cS^{rstI}$ and $\cP^{rstI}$, we can take
$\alpha_{rst}$ and $\beta_{rst}$ to be antisymmetric, and they should
therefore satisfy
\be
(\alpha_{rst}\beta_{pqu}+\beta_{rst}\alpha_{pqu})\varepsilon^{rstpqu}=0\ ,
\label{epcon}
\ee
where there is no summation over $r, s, \cdots, u$.

In the special case where we solve (\ref{epcon}) by taking
\bea
\alpha_{3pq}&=&\cos\omega\ ,\qquad \beta_{3pq}=\sin\omega\ ,
\quad 4\le p\le q\le8\ ,
\nn\\
\alpha_{pqr}&=&\cos\omega\ ,\qquad
\beta_{pqr}= -\sin\omega\ ,\quad 4\le p\le q\le r\le8\ ,
\eea
the boundary conditions become those that would follow from the
${\cal N}=3$ boundary conditions we derived earlier, by decomposing
the ${\cal N}=3$ triplet index $I$ into an ${\cal N}=2$ doublet and a singlet.

If $\omega=0$, i.e. for the de Wit-Nicolai theory, we find that in addition
to the ${\cal N}=2$ supersymmetry-preserving boundary conditions derived
above, there can also
exist another new set of allowed boundary conditions, in which
$A^{rs}_i$ satisfies Neumann boundary conditions. The full set of boundary
conditions in this case takes the form
\bea
&&A^{rI}_{(0)i}\ ,\quad A^{IJ}_{(0)i}=0\ ,\quad A^{rs}_{(1)i}=0\ ,\quad {\chi}^{rIJ}_+=0\ ,
\quad {\chi}^{rsI}_-=0\ ,\quad{\chi}^{rst}_+=0\ ,
\nn\w2
&&{\cal P}^{rsIJ}_{(2)} =0\ ,\quad {\cal S}^{rsIJ}_{(1)}=0\ ,\quad {\cal P}^{rstI}_{(1)}=0\ ,
\quad {\cal S}^{rstI}_{(2)}=0\ .
\eea

\subsection{ $\cN= 1$ Supersymmetry}

In the $\cN=1$ case, we set all the supersymmetry transformation parameters
to zero except $\epsilon^1$. The $SO(8)$ index $I$ is split as
$I = (1, r)$ where $r=2,...,8$.  The supersymmetry variation of the
Dirichlet boundary conditions on the gravitino implies that
\be
A^{r1}_{(0)i}=0\ ,
\ee
which will not impose any further condition on the spin-$\ft12$ fields
because the variation of $A^{r1}_{(0)}$ automatically vanishes under
${\cal N}=1$ supersymmetry. The determination of the boundary conditions
for the remaining fields requires the utilization of the supersymmetry
variations of the leading terms in the Fefferman-Graham expansions for
the spin-0, spin-$1/2$ and spin-1 fields, which are given by
\bea
\delta{\chi}^{1rs}_+ &=& 0\ ,  \qquad\delta{\chi}^{1rs}_-=0\ ,
\nn\w2
\delta{\chi}^{rst}_{+} &=&-{\cal S}^{rst1}_{(2)}{\epsilon}^1_+
+2{\rm i}{\cal P}^{rst1}_{(1)}\gamma_5{\epsilon}^1_--{\rm i}\slashed{D}{\cal P}_{(1)}^{rst1}\gamma_5{\epsilon}^1_+\ ,
\nn\w2
\delta{\chi}^{rst}_{-} &=& 2{\cal S}^{rst1}_{(1)}{\epsilon}^1_--{\rm i}{\cal P}^{rst1}_{(2)}\gamma_5{\epsilon}^1_+
+\slashed{D}{\cal S}_{(1)}^{rst1}{\epsilon}^1_+\ ,
\nn\w2
\delta{\cal S}^{1rst}_{(1)} &=&{\epsilon}^1_+{\chi}^{rst}_-,\quad \delta{\cal P}^{1rst}_{(1)}
=-\im {\epsilon}^1_+\gamma_5{\chi}^{rst}_+\ ,
\nn\w2
\delta{\cal S}^{1rst}_{(2)} &=&\Big(\bar{{\epsilon}}^{1}_-{\chi}_{+}^{rst}
+\bar{{\epsilon}}^{1}_+\slashed{D}{\chi}_{+}^{rst}\Big)\ ,
\nn\w2
\delta{\cal P}^{1rst}_{(2)} &=&-\im \Big(\bar{{\epsilon}}^{1}_-\gamma_5{\chi}_{-}^{rst}
-\bar{{\epsilon}}^{1}_+\gamma_5\slashed{D}{\chi}_{-}^{rst}\Big)\ ,
\nn\w2
\delta A^{rs}_{(0)i} &=&-\Big(\cos2\omega\ {\epsilon}^{1}_+\gamma_{(0)i}{\chi}^{rs1}_{+}
+{\rm i}\sin2\omega\ {\epsilon}^{1}_+\gamma_{(0)i}\gamma_5{\chi}^{rs1}_{-}\Big)\
\nn\w2
\delta A^{rs}_{(1)i} &=&-\Big({\cal S}_{(1)}^{rstp}{\epsilon}^{1}_+\gamma_{(0)i}{\chi}^{tp1}_{+}
+{\rm i}{\cal P}_{(1)}^{rstp}{\epsilon}^{1}_+\gamma_{(0)i}\gamma_5{\chi}^{tp1}_{-}
\nn\\
&&-D_{i}(A_{(0)})(\cos2\omega\ {\epsilon}^1_{+}{\chi}^{rs1}_{-}+{\rm i}\sin2\omega\ {\epsilon}^1_{+}\gamma_5{\chi}^{rs1}_+)\Big)\ ,
\eea
where we have omitted the variations of ${\cal S}^{rstp}$ and
${\cal P}^{rstp}$, since they are dual to ${\cal S}^{rst1}$ and
${\cal P}^{rst1}$ respectively, and therefore are not independent fields.
One can check that there are two sets of boundary conditions preserving
${\cal N}=1$ supersymmetry. The first set, for which the vector fields all
obey Dirichlet boundary conditions, is given by
\bea
&& A^{rs}_{(0)i}=0,\quad\cos2\omega\  {\chi}^{1rs}_++\im \sin2\omega\ \gamma_5{\chi}^{1rs}_-=0,\quad\alpha_{rst}{\chi}^{rst}_++\im\beta_{rst} \gamma_5{\chi}^{rst}_-=0\ ,
\nn\w2
&&\alpha_{rst}{\cal S}^{rst1}_{(2)}-\beta_{rst}{\cal P}^{rst1}_{(2)}=0,\quad \beta_{rst}{\cal S}^{rst1}_{(1)}+\alpha_{rst}{\cal P}^{rst1}_{(1)}=0\ ,
\eea
where the totally-antisymmetric coefficients $\alpha_{rst}$ and
$\beta_{rst}$ are arbitrary constants, and the indices $r,s,t$ are not
summed over. The second set, for which the subset of vectors $A_i^{rs}$
instead satisfy Neumann boundary conditions, takes the form
\bea
&&A^{rs}_{(1)i}=0\ ,\quad \sin2\omega\  {\chi}^{1rs}_+-\im \cos2\omega\ \gamma_5{\chi}^{1rs}_-=0,\quad\sin2\omega\ {\chi}^{rst}_+
-\im\cos2\omega\  \gamma_5{\chi}^{rst}_-=0\ ,\nn
\w2
&&\sin2\omega\ {\cal S}^{rst1}_{(2)} +\cos2\omega\ {\cal P}^{rst1}_{(2)}=0\ ,\quad
\cos2\omega\ {\cal S}^{rst1}_{(1)}-\sin2\omega\ {\cal P}^{rst1}_{(1)}=0\ .
\eea

     We now study the possibility of imposing mixed boundary
conditions (\ref{mbc}) on $A^{rs}_i$, of the form
\be
\alpha A_{(1)i}^{rs} + \beta \varepsilon_{ijk} F_{(0)jk}^{rs}=0\ .
\label{mbc}
\ee
Upon using (see, for example, \cite{lupoto})),
\be
\partial_i\epsilon_+=\ft{1}{2}\gamma_i\epsilon_-\ ,
\label{KS}
\ee
the supersymmetry variation of (\ref{mbc}) gives rise to terms
proportional to $\epsilon_+$ and $\epsilon_-$. The vanishing of the terms
proportional to $\epsilon_-$ requires
\be
\chi^{rs1}_-=\im \nu \gamma_5 \chi^{rs1}_+,\qquad \nu=\frac{\alpha\sin2\omega+\beta\cos2\omega}{\alpha\cos2\omega-\beta\sin2\omega}\ .
\label{nu}
\ee
The vanishing of terms proportional to $\epsilon_+$, on the other hand,
requires
\bea
&&\alpha({\cS}^{rstp}-\nu{\cP}^{rstp})\gamma_i\chi_+^{tp1}+\im \alpha(\nu\cos2\omega+\sin2\omega)\gamma_5D_i(A_0)\chi_+^{rs1}\nn \w2
&&+\beta\varepsilon^{ijk}(\cos2\omega-\nu\sin2\omega)\gamma_kD_{j}(A_0)\chi_+^{rs1}=0\ .
\eea
The coefficient of the first group of terms can be made to vanish. However,
the vanishing of last two groups of terms independently leads to the
conditions $\omega=0$ and $\nu=0$. From the definition of $\nu$ in
(\ref{nu}), it follows that $\beta=0$. Thus we conclude that mixed boundary
conditions are not allowed by $\cN=1$ supersymmetry.

The terms proportional to $D_i\chi_+$ play the key role in ruling out
the mixed boundary conditions. This makes it straightforward to show
that mixed boundary conditions of the form (\ref{mbc}) are also forbidden
by ${\cN}>1$ supersymmetry. This is due to the fact that the $D_i\chi_+$
terms from $\delta A^{rs}_{(0)i}$ and $\delta A^{rs}_{(1)i}$ are universal,
with the appropriate ranges of the indices $(I,r)$ understood.  Thus our
previous assumptions of no mixed boundary conditions in the ${\cal N}\ge 2$
analyses were justified.

\end{appendix}

\newpage


\end{document}